\pdfoutput=1
\documentclass[12pt,a4paper]{article}

\usepackage{ifthen} 
\newboolean{pdflatex}
\setboolean{pdflatex}{true} 

\newboolean{articletitles}
\setboolean{articletitles}{true} 

\newboolean{uprightparticles}
\setboolean{uprightparticles}{true} 

\def\paperauthors{LHCb collaboration} 
\def\paperasciititle{Observation of Lb -->D+ p pi- pi-  
and  Lb -->D*+ p pi- pi-  
decays}  
\def\papertitle{Observation of 
\mbox{$\decay{\Lb}{\Dp\proton\pim\pim}$} and 
\mbox{$\decay{\Lb}{\Dstarp\proton\pim\pim}$}~decays} 

\def\paperkeywords{{High Energy Physics}, {LHCb}} 
\def\papercopyright{\the\year\ CERN for the benefit of the LHCb collaboration} 
\def\paperlicence{CC BY 4.0 licence}
\def\paperlicenceurl{https://creativecommons.org/licenses/by/4.0/}

\usepackage[top=1in, bottom=1.25in, left=1in, right=1in]{geometry}

\usepackage{microtype}
\usepackage{lineno}  
\usepackage{xspace} 
\usepackage{caption} 

\usepackage{graphicx}  
\usepackage[dvipsnames]{xcolor}
\usepackage{colortbl}
\graphicspath{{./figs/}} 

\usepackage{amsmath} 
\usepackage{amssymb}
\usepackage{amsfonts}
\usepackage{upgreek} 

\newcommand*\patchAmsMathEnvironmentForLineno[1]{%
\expandafter\let\csname old#1\expandafter\endcsname\csname #1\endcsname
\expandafter\let\csname oldend#1\expandafter\endcsname\csname
end#1\endcsname
 \renewenvironment{#1}%
   {\linenomath\csname old#1\endcsname}%
   {\csname oldend#1\endcsname\endlinenomath}%
}
\newcommand*\patchBothAmsMathEnvironmentsForLineno[1]{%
  \patchAmsMathEnvironmentForLineno{#1}%
  \patchAmsMathEnvironmentForLineno{#1*}%
}
\AtBeginDocument{%
\patchBothAmsMathEnvironmentsForLineno{equation}%
\patchBothAmsMathEnvironmentsForLineno{align}%
\patchBothAmsMathEnvironmentsForLineno{flalign}%
\patchBothAmsMathEnvironmentsForLineno{alignat}%
\patchBothAmsMathEnvironmentsForLineno{gather}%
\patchBothAmsMathEnvironmentsForLineno{multline}%
\patchBothAmsMathEnvironmentsForLineno{eqnarray}%
}

\usepackage{hyperxmp}

\usepackage[pdftex,
            pdfauthor={\paperauthors},
            pdftitle={\paperasciititle},
            pdfkeywords={\paperkeywords},
            pdfcopyright={Copyright (C) \papercopyright},
            pdflicenseurl={\paperlicenceurl}]{hyperref}

\usepackage[colorinlistoftodos,textsize=scriptsize]{todonotes}

\usepackage[bottom,flushmargin,hang,multiple]{footmisc}

\usepackage[all]{hypcap} 

\usepackage{xspace} 
\usepackage{upgreek}

\def\lhcb   {\mbox{LHCb}\xspace}

\def\MagUp {\mbox{\em Mag\kern -0.05em Up}\xspace}

\ifthenelse{\boolean{uprightparticles}}%
{
 
 \def\Pgamma      {\ensuremath{\upgamma}\xspace}

 \def\Pnu         {\ensuremath{\upnu}\xspace}                 
                  
 \def\Ppi         {\ensuremath{\uppi}\xspace}                 
                  
 \def\Prho        {\ensuremath{\uprho}\xspace}

 \def\Pphi        {\ensuremath{\upphi}\xspace}

 \def\PDelta      {\ensuremath{\Delta}\xspace}                 
 \def\PXi         {\ensuremath{\Xi}\xspace}                 
 \def\PLambda     {\ensuremath{\Lambda}\xspace}                 
 \def\PSigma      {\ensuremath{\Sigma}\xspace}                 
 \def\POmega      {\ensuremath{\Omega}\xspace}                 
 \def\PUpsilon    {\ensuremath{\Upsilon}\xspace}

 \def\PB      {\ensuremath{\mathrm{B}}\xspace}                 
                  
 \def\PD      {\ensuremath{\mathrm{D}}\xspace}

 \def\PK      {\ensuremath{\mathrm{K}}\xspace}

 \def\Pa      {\ensuremath{\mathrm{a}}\xspace}                 
 \def\Pb      {\ensuremath{\mathrm{b}}\xspace}                 
 \def\Pc      {\ensuremath{\mathrm{c}}\xspace}                 
 \def\Pd      {\ensuremath{\mathrm{d}}\xspace}                 
 \def\Pe      {\ensuremath{\mathrm{e}}\xspace}                 
 \def\Pf      {\ensuremath{\mathrm{f}}\xspace}

 \def\Pi      {\ensuremath{\mathrm{i}}\xspace}

 \def\Pp      {\ensuremath{\mathrm{p}}\xspace}

 \def\Ps      {\ensuremath{\mathrm{s}}\xspace}                 
                  
 \def\Pu      {\ensuremath{\mathrm{u}}\xspace}

 \def\thebaroffset{0.0em}
}
{
 
 \def\Pgamma      {\ensuremath{\gamma}\xspace}

 \def\Pnu         {\ensuremath{\nu}\xspace}                 
                  
 \def\Ppi         {\ensuremath{\pi}\xspace}                 
                  
 \def\Prho        {\ensuremath{\rho}\xspace}

 \def\Pphi        {\ensuremath{\phi}\xspace}

 \mathchardef\PDelta="7101
 \mathchardef\PXi="7104
 \mathchardef\PLambda="7103
 \mathchardef\PSigma="7106
 \mathchardef\POmega="710A
 \mathchardef\PUpsilon="7107
                  
 \def\PB      {\ensuremath{B}\xspace}                 
                  
 \def\PD      {\ensuremath{D}\xspace}

 \def\PK      {\ensuremath{K}\xspace}

 \def\Pa      {\ensuremath{a}\xspace}                 
 \def\Pb      {\ensuremath{b}\xspace}                 
 \def\Pc      {\ensuremath{c}\xspace}                 
 \def\Pd      {\ensuremath{d}\xspace}                 
 \def\Pe      {\ensuremath{e}\xspace}                 
 \def\Pf      {\ensuremath{f}\xspace}

 \def\Pi      {\ensuremath{i}\xspace}

 \def\Pp      {\ensuremath{p}\xspace}

 \def\Ps      {\ensuremath{s}\xspace}                 
                  
 \def\Pu      {\ensuremath{u}\xspace}

 \def\thebaroffset{0.18em}
}
\newcommand{\offsetoverline}[2][\thebaroffset]{\kern #1\overline{\kern -#1 #2}}%

\makeatletter
\ifcase \@ptsize \relax
  \newcommand{\miniscule}{\@setfontsize\miniscule{4}{5}}
\or
  \newcommand{\miniscule}{\@setfontsize\miniscule{5}{6}}
\or
  \newcommand{\miniscule}{\@setfontsize\miniscule{5}{6}}
\fi
\makeatother

\DeclareRobustCommand{\optbar}[1]{\shortstack{{\miniscule (\rule[.5ex]{1.25em}{.18mm})}
  \\ [-.7ex] $#1$}}

\def\epem       {{\ensuremath{\Pe^+\Pe^-}}\xspace}

\def\g      {{\ensuremath{\Pgamma}}\xspace}

\def\uquark    {{\ensuremath{\Pu}}\xspace}
\def\uquarkbar {{\ensuremath{\overline \uquark}}\xspace}

\def\dquark    {{\ensuremath{\Pd}}\xspace}

\def\squark    {{\ensuremath{\Ps}}\xspace}

\def\cquark    {{\ensuremath{\Pc}}\xspace}
\def\cquarkbar {{\ensuremath{\overline \cquark}}\xspace}

\def\bquark    {{\ensuremath{\Pb}}\xspace}

\def\pion   {{\ensuremath{\Ppi}}\xspace}
\def\piz    {{\ensuremath{\pion^0}}\xspace}
\def\pip    {{\ensuremath{\pion^+}}\xspace}
\def\pim    {{\ensuremath{\pion^-}}\xspace}
\def\pipm   {{\ensuremath{\pion^\pm}}\xspace}

\def\rhomeson {{\ensuremath{\Prho}}\xspace}

\def\rhom     {{\ensuremath{\rhomeson^-}}\xspace}

\def\kaon    {{\ensuremath{\PK}}\xspace}

\def\KorKbar {\kern \thebaroffset\optbar{\kern -\thebaroffset \PK}{}\xspace}

\def\Kp      {{\ensuremath{\kaon^+}}\xspace}
\def\Km      {{\ensuremath{\kaon^-}}\xspace}

\def\Kmp     {{\ensuremath{\kaon^\mp}}\xspace}
\def\KS      {{\ensuremath{\kaon^0_{\mathrm{S}}}}\xspace}

\def\Dbar    {{\ensuremath{\offsetoverline{\PD}}}\xspace}
\def\D       {{\ensuremath{\PD}}\xspace}

\def\DorDbar {\kern \thebaroffset\optbar{\kern -\thebaroffset \PD}\xspace}
\def\Dz      {{\ensuremath{\D^0}}\xspace}
\def\Dzb     {{\ensuremath{\Dbar{}^0}}\xspace}
\def\Dp      {{\ensuremath{\D^+}}\xspace}
\def\Dm      {{\ensuremath{\D^-}}\xspace}

\def\DpDm    {\ensuremath{\Dp {\kern -0.16em \Dm}}\xspace}
\def\Dstar   {{\ensuremath{\D^*}}\xspace}

\def\Dstarp  {{\ensuremath{\D^{*+}}}\xspace}

\def\Ds      {{\ensuremath{\D^+_\squark}}\xspace}

\def\Dsm     {{\ensuremath{\D^-_\squark}}\xspace}

\def\B       {{\ensuremath{\PB}}\xspace}

\def\BorBbar {\kern \thebaroffset\optbar{\kern -\thebaroffset \PB}\xspace}

\def\Bd      {{\ensuremath{\B^0}}\xspace}

\def\BdorBdbar {\kern \thebaroffset\optbar{\kern -\thebaroffset \Bd}\xspace}

\def\Bs      {{\ensuremath{\B^0_\squark}}\xspace}

\def\BsorBsbar {\kern \thebaroffset\optbar{\kern -\thebaroffset \Bs}\xspace}

\def\Y#1S{\ensuremath{\PUpsilon{(#1S)}}\xspace}

\def\proton      {{\ensuremath{\Pp}}\xspace}

\def\Lz          {{\ensuremath{\PLambda}}\xspace}

\def\LorLbar     {\kern \thebaroffset\optbar{\kern -\thebaroffset \PLambda}\xspace}

\def\Xires       {{\ensuremath{\PXi}}\xspace}

\def\Lc          {{\ensuremath{\Lz^+_\cquark}}\xspace}

\def\Lb           {{\ensuremath{\Lz^0_\bquark}}\xspace}

\def\Xibz         {{\ensuremath{\Xires^0_\bquark}}\xspace}

\def\BF         {{\ensuremath{\mathcal{B}}}\xspace}
\def\BR         {\BF}

\newcommand{\decay}[2]{\ensuremath{#1\!\to #2}\xspace} 

\def\to                 {\ensuremath{\rightarrow}\xspace}

\def\CP                {{\ensuremath{C\!P}}\xspace}

\def\AT#1     {\ensuremath{A_{\mathrm{T}}^{#1}}\xspace}           

\def\C#1      {\ensuremath{\mathcal{C}_{#1}}\xspace}                       
\def\Cp#1     {\ensuremath{\mathcal{C}_{#1}^{'}}\xspace}                    
\def\Ceff#1   {\ensuremath{\mathcal{C}_{#1}^{\mathrm{(eff)}}}\xspace}        
\def\Cpeff#1  {\ensuremath{\mathcal{C}_{#1}^{'\mathrm{(eff)}}}\xspace}       
\def\Ope#1    {\ensuremath{\mathcal{O}_{#1}}\xspace}                       
\def\Opep#1   {\ensuremath{\mathcal{O}_{#1}^{'}}\xspace}                    

\newcommand{\nospaceunit}[1]{\ensuremath{\text{#1}}}       
\newcommand{\aunit}[1]{\ensuremath{\text{\,#1}}}       

\newcommand{\tev}{\aunit{Te\kern -0.1em V}\xspace}
\newcommand{\gev}{\aunit{Ge\kern -0.1em V}\xspace}
\newcommand{\mev}{\aunit{Me\kern -0.1em V}\xspace}
\newcommand{\kev}{\aunit{ke\kern -0.1em V}\xspace}
\newcommand{\ev}{\aunit{e\kern -0.1em V}\xspace}
 
\newcommand{\mevc}{\ensuremath{\aunit{Me\kern -0.1em V\!/}c}\xspace}
\newcommand{\gevc}{\ensuremath{\aunit{Ge\kern -0.1em V\!/}c}\xspace}
\newcommand{\mevcc}{\ensuremath{\aunit{Me\kern -0.1em V\!/}c^2}\xspace}
\newcommand{\gevcc}{\ensuremath{\aunit{Ge\kern -0.1em V\!/}c^2}\xspace}

\def\mum  {\ensuremath{\,\upmu\nospaceunit{m}}\xspace}

\def\fb   {\ensuremath{\aunit{fb}}\xspace}
\def\invfb   {\ensuremath{\fb^{-1}}\xspace}

\newcommand{\chisq}{\ensuremath{\chi^2}\xspace}

\newcommand{\chisqip}{\ensuremath{\chi^2_{\text{IP}}}\xspace}

\def\gsim{{~\raise.15em\hbox{$>$}\kern-.85em
          \lower.35em\hbox{$\sim$}~}\xspace}
\def\lsim{{~\raise.15em\hbox{$<$}\kern-.85em
          \lower.35em\hbox{$\sim$}~}\xspace}

\def\sPlot{\mbox{\em sPlot}\xspace}

\def\sqs   {\ensuremath{\protect\sqrt{s}}\xspace}

\def\pt         {\ensuremath{p_{\mathrm{T}}}\xspace}

\def\ptot       {\ensuremath{p}\xspace}

\def\evtgen     {\mbox{\textsc{EvtGen}}\xspace}

\def\geant      {\mbox{\textsc{Geant4}}\xspace}

\def\photos     {\mbox{\textsc{Photos}}\xspace}

\def\pythia     {\mbox{\textsc{Pythia}}\xspace}

\def\tell1  {TELL1\xspace}
\def\ukl1   {UKL1\xspace}

\newcommand{\eg}{\mbox{\itshape e.g.}\xspace}
\newcommand{\ie}{\mbox{\itshape i.e.}\xspace}

\usepackage{cite} 
\usepackage{mciteplus}

\usepackage{longtable} 

\usepackage{graphpap} 
\usepackage{multirow} 
\usepackage{rotating} 
\usepackage{afterpage}
\usepackage{dashrule}
\usepackage{wasysym}
\usepackage{cancel}

\newcommand{\kevc}{\ensuremath{\aunit{ke\kern -0.1em V\!/}c}\xspace}
\newcommand{\kevcc}{\ensuremath{\aunit{ke\kern -0.1em V\!/}c^2}\xspace}

\usepackage{scalerel}

\def\XXint#1#2#3{{\setbox0=\hbox{$#1{#2#3}{\int}$}
     \vcenter{\hbox{$#2#3$}}\kern-.5\wd0}}

\usepackage[normalem]{ulem}

\makeatletter
\g@addto@macro\bfseries{\boldmath}
\makeatother

\begin{document}

\renewcommand{\thefootnote}{\fnsymbol{footnote}}
\setcounter{footnote}{1}


\begin{titlepage}
\pagenumbering{roman}

\vspace*{-1.5cm}
\centerline{\large EUROPEAN ORGANIZATION FOR NUCLEAR RESEARCH (CERN)}
\vspace*{1.5cm}
\noindent
\begin{tabular*}{\linewidth}{lc@{\extracolsep{\fill}}r@{\extracolsep{0pt}}}
\ifthenelse{\boolean{pdflatex}}
{\vspace*{-1.5cm}\mbox{\!\!\!\includegraphics[width=.14\textwidth]{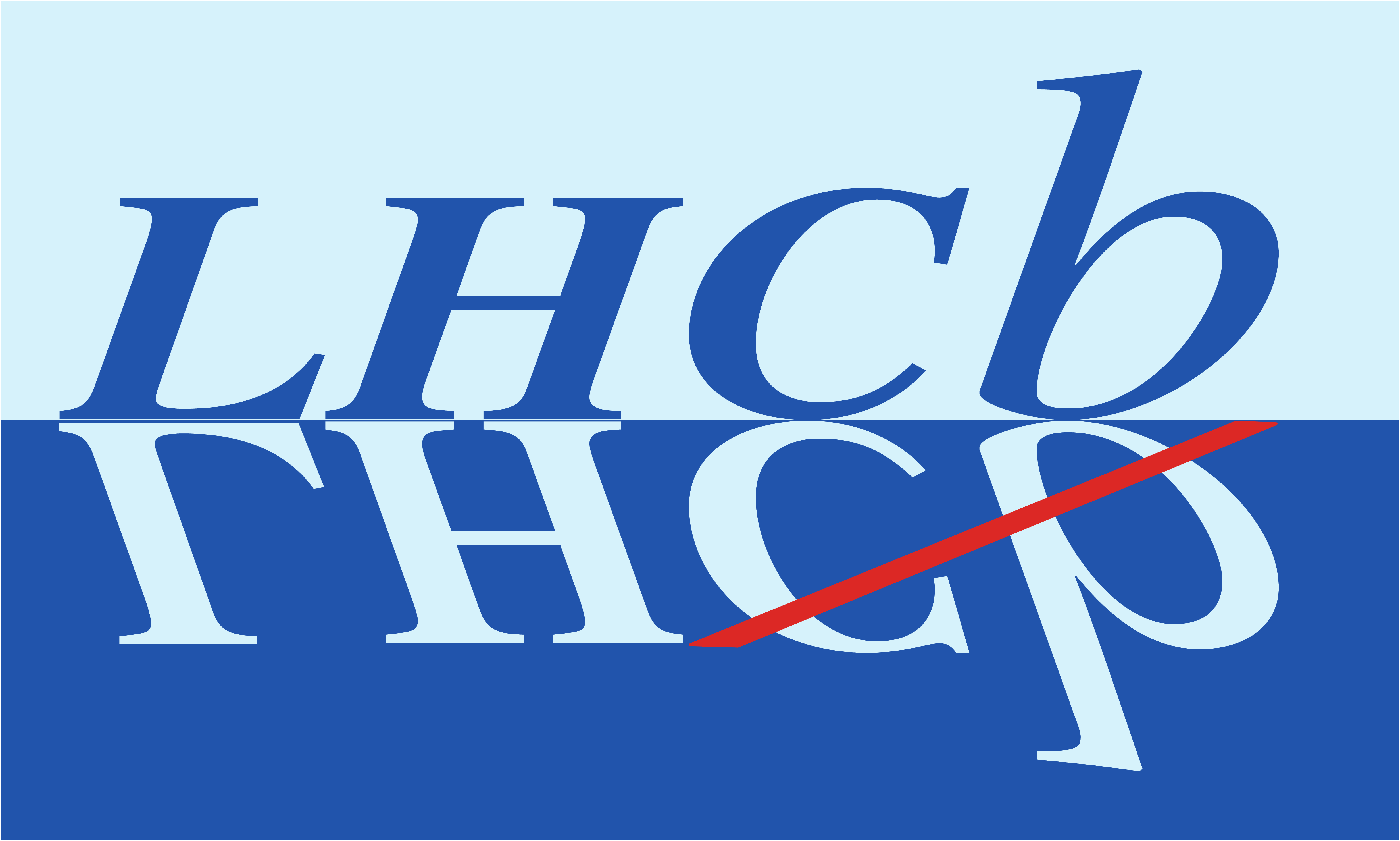}} & &}%
{\vspace*{-1.2cm}\mbox{\!\!\!\includegraphics[width=.12\textwidth]{figs/lhcb-logo.eps}} & &}%
\\
 & & CERN-EP-2021-237 \\  
 & & LHCb-PAPER-2021-040 \\  
 & & March 25, 2021 \\ 
 & & \\
%
\end{tabular*}

\vspace*{2.0cm}

{\normalfont\bfseries\boldmath\huge
\begin{center}
  \papertitle 
\end{center}
}

\vspace*{1.0cm}

\begin{center}
\paperauthors\footnote{Authors are listed at the end of this paper.}
\end{center}

\vspace{\fill}

\begin{abstract}
The~multihadron decays \mbox{$\decay{\Lb}{\Dp\proton\pim\pim}$} 
and \mbox{$\decay{\Lb}{\Dstarp\proton\pim\pim}$} 
are observed in data corresponding to an~integrated luminosity of 3\invfb, collected in proton\nobreakdash-proton collisions at 
centre\nobreakdash-of\nobreakdash-mass energies of 7 and 8\tev by the~\lhcb~detector.
Using the~decay \mbox{$\decay{\Lb}{\Lc\pip\pim\pim}$}
as a~normalisation channel, 
the~ratio of branching fractions is measured to be 
\begin{equation*}
        \dfrac { \BR({\decay{\Lb}{\Dp\proton\pim\pim}})}
               { \BR({\decay{\Lb}{\Lc\pip\pim\pim}})}
        \times
        \dfrac{ \BR({\decay{\Dp}{\Km\pip\pip}})} 
              { \BR({\decay{\Lc}{\proton\Km\pip}})} 
        =  \left( 5.35 \pm 0.21 \pm 0.16 \right) \% \,,
\end{equation*}
where the first uncertainty is statistical 
and the~second systematic. 
The~ratio of branching fractions for 
the~ \mbox{$\decay{\Lb}{\Dstarp\proton\pim\pim}$} 
and \mbox{$\decay{\Lb}{\Dp\proton\pim\pim}$}~decays 
is found to be 
\begin{equation*}
        {\dfrac{\BR({\decay{\Lb}{\Dstarp\proton\pim\pim}})}
           {\BR({\decay{\Lb}{\Dp\proton\pim\pim}})}} \times
           \left( 
           \BR({\decay{\Dstarp}{\Dp\piz}}) + 
           \BR({\decay{\Dstarp}{\Dp\g}}) 
           \right) 
          = \left( 61.3 \pm 4.3 \pm 4.0  \right) \% \,.
  \end{equation*}
\end{abstract}

\vspace*{2.0cm}

\begin{center}
Published in JHEP 03 (2022) 153
\end{center}

\vspace{\fill}

{\footnotesize 
\centerline{\copyright~\papercopyright. \href{\paperlicenceurl}{\paperlicence}.}}
\vspace*{2mm}

\end{titlepage}


\newpage
\setcounter{page}{2}
\mbox{~}

\cleardoublepage
%
%
%
%
%


\renewcommand{\thefootnote}{\arabic{footnote}}
\setcounter{footnote}{0}

\cleardoublepage


\pagestyle{plain} 
\setcounter{page}{1}
\pagenumbering{arabic}




\clearpage
\section{Introduction}
\label{sec:Introduction}
  Nonleptonic decays with multiple hadrons,
  such as 
 \mbox{$\decay{\Lb}{\Dp\proton\pim\pim}$}  and
 \mbox{$\decay{\Lb}{\Lc\pim\pip\pim}$},  
 are a
  useful platform for testing non\nobreakdash-perturbative 
  quantum chromodynamics (QCD) approaches such as QCD factorisation (QCDF). 
  At~the~quark level these \Lb baryon decays 
  are mediated by the~weak 
\mbox{$\decay{\bquark}{\cquark\cquarkbar\squark}$} and 
\mbox{$\decay{\bquark}{\cquark\uquarkbar\dquark}$} 
transitions.\footnote{The inclusion 
  of charge-conjugate processes is implied throughout the paper.}
Calculating the~rates for these decays  is more challenging than for their semileptonic 
 \mbox{$\decay{\bquark}{\cquark\ell^-\bar{\Pnu}_{\ell}}$}
 partners, since  
 strong interactions are present in both 
 the~hadronic initial and final states.  
 Despite these difficulties, 
 which are  due to QCD effects,  
 substantial progress has been made in computing 
 hadronic two\nobreakdash-body and 
 quasi\nobreakdash-two\nobreakdash-body decays;
 earlier calculations~\cite{Mannel:1992ti,
 Guo:1998at,
 Fayyazuddin:1998aa,
 Leibovich:2003tw} have  
 been refined in Refs.~\cite{Huber:2016xod,
 Gutsche:2018utw}. 
  These theory predictions
  agree well with both the CDF measurement of \Lb~production and decays~\cite{Abulencia:2006df}, 
  and a similar LHCb measurement~\cite{LHCb-PAPER-2014-004}.
  Formulated within the framework of QCDF,
    these predictions
  are calculated for several
  decay modes, 
  \mbox{$\decay{\Lb}{\Lc(\pim,\rhom,\Pa_{1}^{-})}$}, including exclusive modes where the~intermediate 
  resonance decays 
  into a~final state with multiple pions,
  \eg 
  \mbox{$\decay{\Pa_{1}^{-}}{\pim\pim\pip}$}~\cite{Huber:2016xod}. 
  Such decay channels contribute to 
  the~multihadron \(\Lb\to\Lc\pim\pip\pim \) 
  decay analysed in this study. 

 Final state protons and charm mesons 
 are of particular interest in 
 multihadron decays of beauty baryons, 
 where the~\cquark-quark from 
 the~$\bquark\to\cquark$~transition 
 hadronises into the final state 
 separate from the~baryon,
 \ie a~charm meson and a~proton.
 This topology is not only important for 
 charm baryon and meson spectroscopy, 
 but also sensitive to QCD effects in beauty 
 baryons as well as charm-quark hadronisation. However, 
 this topology has not been widely studied. 
  Currently, only a~few decay modes of beauty baryons with 
  the final state configuration described 
  above are known:
  \mbox{$\decay{\Lb}{\Dz\proton\pim}$}~\cite{Basile:1981wr,
  Bari:1991ty,
  LHCb-PAPER-2013-056},
  \mbox{$\decay{\Lb}{\Dz\proton\Km}$}
  and 
  \mbox{$\decay{\Xibz}{\Dz\proton\Km}$}\cite{LHCb-PAPER-2013-056}.
  The~amplitude analysis of
  \mbox{$\decay{\Lb}{\Dz\proton\Km}$}~decays 
  discovered a~rich resonance structure 
  allowing the~study of excited 
  charm baryons~\cite{LHCb-PAPER-2016-061}. 
  Recently, the~\lhcb collaboration reported an observation of 
  the~\mbox{$\decay{\Lb}{ \D \proton \Km}$} channel with 
  a~\mbox{$\decay{\D}{\Kmp \pipm}$} decay, where the~state 
  $\D$ 
  is a~superposition of $\Dz$ and 
  $\Dzb$~states~\cite{LHCb-PAPER-2021-027}.
  The~\CP asymmetry 
  in this decay 
  and the~ratio of branching fractions for 
  the~\mbox{$\decay{\Lb}{  \left( \decay{\D}
  {\Km\pip} \right) \proton \Km}$}
  and \mbox{$\decay{\Lb}{ \left( \decay{\D}{\Kp\pim} \right) 
  \proton \Km}$}~decays
  are also measured.

In this paper, the~first observation of 
the~\mbox{$\decay{\Lb}{\Dp\proton\pim\pim}$}
  and \mbox{$\decay{\Lb}{\Dstarp\proton\pim\pim}$} 
  multihadron decay modes is reported.
  The~measurements are based on proton-proton (\proton\proton)~collision data, 
  corresponding to integrated luminosities 
  of \(1\) and \(2\invfb\) collected with the LHCb detector
  at center-of-mass energies of \(7\) and \(8\tev \), respectively.
  The following ratios of branching fractions are reported 
  \begin{subequations} \label{eq:R}
  \begin{eqnarray}
    {\mathcal{R}}_{\Dp} 
    & \equiv  & 
        \dfrac { \BR({\decay{\Lb}{\Dp\proton\pim\pim}})}
               { \BR({\decay{\Lb}{\Lc\pip\pim\pim}})}
        \times
        \dfrac{ \BR({\decay{\Dp}{\Km\pip\pip}})} 
              { \BR({\decay{\Lc}{\proton\Km\pip}})} 
        \label{eq:rdp}    \,,   
        \\
        {\mathcal{R}}_{\Dstarp}
        & \equiv & 
        {\dfrac{\BR({\decay{\Lb}{\Dstarp\proton\pim\pim})}}
           {\BR({\decay{\Lb}{\Dp\proton\pim\pim}})}} \times 
           \BR({\decay{\Dstarp}{\Dp\piz/\g}}) \,, \label{eq:rdstarp} 
        \end{eqnarray}
  \end{subequations}
 where $\BR({\decay{\Dstarp}{\Dp\piz/\g}})$ equals  
 $\BR({\decay{\Dstarp}{\Dp\piz}}) + 
 \BR({\decay{\Dstarp}{\Dp\g}})$,  
 and 
 the~\mbox{$\decay{\Lb}{\Lc\pip\pim\pim}$}~mode
 with the~\mbox{$\decay{\Lc}{\proton\Km\pip}$} decay
 is used as the normalisation channel. 
 No theory predictions are currently available for  
 the decay modes \mbox{$\decay{\Lb}{\Dp\proton\pim\pim}$} 
 and \mbox{$\decay{\Lb}{\Dstarp\proton\pim\pim}$}.

\section{Detector and simulation}
\label{sec:Detector}
  The \lhcb detector~\cite{LHCb-DP-2008-001,LHCb-DP-2014-002} 
  is a~single-arm forward spectrometer
  covering the~\mbox{pseudorapidity} 
  range \(2<\eta <5\), designed for 
  the~study of particles containing
  \bquark or \cquark quarks. 
  The~detector includes a~high-precision tracking 
  system consisting of 
  a~silicon\nobreakdash-strip vertex detector 
  surrounding the~\proton\proton~interaction region, 
  a~large-area silicon-strip
  detector located upstream 
  of a dipole magnet with a bending power of about \(4{\mathrm{\,Tm}}\), 
  and
  silicon-strip detectors and straw drift tubes placed
  downstream of the~magnet.
  The~tracking system provides 
  a~measurement of the~momentum, \ptot, of charged particles 
  with a~relative uncertainty that varies from 0.5\% at low 
  momentum to 1.0\% at 200\(\gevc\).  
  The~minimum
  distance of a track to 
  a~primary \(\proton\proton\) collision vertex\,(PV), 
  the~impact parameter\,(IP), 
  is measured
  with a resolution of 
  \((15+29/\pt)\mum\), where \pt is the component
  of the~momentum transverse to the
  beam, in\,\gevc.  Different types of charged hadrons are 
  distinguished using information from two
  ring-imaging Cherenkov detectors.  
  Photons, electrons and hadrons are identified by a calorimeter
  system consisting of scintillating-pad and preshower detectors, 
  an~electromagnetic and a~hadronic
  calorimeter. 
  Muons are identified by a~system composed of alternating 
  layers of iron and multiwire
  proportional chambers.
%
%

The online event selection is performed by a trigger system. 
The~trigger consists of a~hardware stage, 
based on information from the~calorimeter and muon
systems, 
followed by a~software stage, 
which applies a~full event reconstruction~\cite{LHCb-DP-2012-004}.
The~events used in this analysis 
are selected at the~hardware stage by requiring 
a~cluster in the~calorimeters with transverse 
energy greater than $3.6\gev$.
The~software trigger requires 
a~two-, three- or four-track secondary vertex with a~large \pt sum of the~particles and 
a~significant displacement 
from the primary $\proton\proton$~interaction 
vertices\,(PVs).
At least one charged particle should have $\pt > 1.7\gevc$ 
and large \chisqip 
with respect to any PV, 
where \chisqip is defined as the
difference in fit \chisq of a~given 
PV reconstructed with and without the considered track.
A multivariate algorithm is used for the identification of 
secondary vertices consistent with 
the~decay of a~\bquark~hadron~\cite{BBDT}. 

  Simulated collision events are used to model the effects of the~detector acceptance 
  and the~imposed selection requirements for signal decay modes.  In the simulation,
  $\proton\proton$~collisions are generated using \pythia~\cite{Sjostrand:2007gs}
  with a specific \lhcb configuration~\cite{LHCb-PROC-2010-056}.
  The~\pt~and rapidity spectra of 
  the~\Lb~baryons in simulation are corrected to match those 
  for the~reconstructed \mbox{$\decay{\Lb}{\Lc\pip\pim\pim}$}~decays,
  which constitute  a~large data 
  sample used for normalisation.
  Decays of unstable particles are described by \evtgen~\cite{Lange:2001uf}, in which 
  final-state radiation is generated using \photos~\cite{Golonka:2005pn}.  
  A~four\nobreakdash-body phase\nobreakdash-space
  decay model is used for  
  the~\mbox{$\decay{\Lb}{\Dp\proton\pim\pim}$}
  and \mbox{$\decay{\Lb}{\Dstarp\proton\pim\pim}$}~decay modes.
  The~decays  \mbox{$\decay{\Lb}{\Lc\pip\pim\pim}$} are simulated
  as a~mixture of 
  decays via intermediate 
  excited $\Sigma^{(*)}_{\cquark}$~resonances
  \mbox{$\decay{\Lb}{ \bigl( \decay{\Sigma_{\cquark}^{(*)++}}{\Lc\pip}
  \bigr) \pim\pim }$} and 
  \mbox{$\decay{\Lb}{ \bigl( \decay{\Sigma_{\cquark}^{(*)0}}{\Lc\pim} \bigr)
  \pip\pim }$};
  excited $\Lc$~baryons 
  \mbox{$\decay{\Lb}{ \bigl(
  \decay{\Lambda_{\cquark}(2595)^+}{\Lc\pip\pim}\bigr)
  \pim }$} and 
  \mbox{$\decay{\Lb}{ \bigl(
  \decay{\Lambda_{\cquark}(2625)^+}{\Lc\pip\pim}\bigr)
  \pim }$}; 
  or light unflavoured hadrons 
  \mbox{$\decay{\Lb}{\Lc\Pa_1^-}$},
  \mbox{$\decay{\Lb}{\Lc\Prho^0\pim}$},
  and \mbox{$\decay{\Lb}{\Lc\Pf_2(1270)\pim}$}.
  The~decay models  are corrected to reproduce 
  the~ten two- and three\nobreakdash-body
  mass distributions from the~signals 
  observed in data.
  The~corrections are applied subsequently 
  for ten mass distributions in several iterations
  untill convergence is achieved. 
  The~interaction of the~generated particles with 
  the~detector and 
  its response are implemented using 
  the~\geant toolkit~\cite{Allison:2006ve, *Agostinelli:2002hh} 
  as described in Ref.~\cite{LHCb-PROC-2011-006}.
To~account for imperfections in the~simulation of
charged\nobreakdash-particle reconstruction, 
the~track reconstruction efficiency
determined from simulation 
is corrected using control channels in data~\cite{LHCb-DP-2013-002}.

\section{Event selection} 
\label{sec:evtsel}

The~\mbox{$\decay{\Lb}{\Dp\proton\pim\pim}$}
and~\mbox{$\decay{\Lb}{\Lc\pip\pim\pim}$}~decays 
are reconstructed using 
the~\mbox{$\decay{\Dp}{\Km\pip\pip}$}
and \mbox{$\decay{\Lc}{\proton\Km\pip}$}~decay channels, 
respectively.
The~selection 
begins with  good\nobreakdash-quality 
reconstructed charged tracks
that are inconsistent 
with being produced in 
a~$\proton\proton$~interaction 
vertex. 
Kaons, pions and protons, identified using information 
from the~RICH detectors~\cite{LHCb-PROC-2011-008,LHCb-DP-2012-003}, are selected 
from well\nobreakdash-reconstructed tracks
within the~acceptance of the~spectrometer 
with $\pt>100\mevc$.
To~allow for efficient  particle identification,
kaons and pions are required 
to have a~momentum between 3~and~120\gevc,
while protons must have momenta between 9 and 120\gevc.

The~\mbox{$\decay{\Dp}{\Km\pip\pip}$} and 
\mbox{$\decay{\Lc}{\proton\Km\pip}$}~candidates 
are reconstructed from selected kaon, pion and proton candidates 
requiring  $\Km\pip\pip$  and $\proton\Km\pip$ 
combinations to 
form a~good quality three\nobreakdash-prong 
common vertex, 
which is significantly separated from any~PV.
A~reconstructed mass 
for the~\Dp and \Lc~candidates
is required to be 
within $\pm34$ and $\pm24\mevcc$~mass 
windows around the~known masses 
of the~\Dp and \Lc~hadrons~\cite{PDG2021}, respectively.
These~mass ranges correspond to
approximately $\pm4\upsigma_m$~regions,   
where $\upsigma_{m}$~is the~mass resolution.
Three-track combinations are also formed of $\proton\pim\pim$ and $\pip\pim\pim$ particle triplets, and are required to 
have a~good\nobreakdash-quality 
common vertex that is distinct from the~PV. 
The~mass of these~$\proton\pim\pim$ and $\pip\pim\pim$~combinations
are required to be below 4 and 3\gevcc, respectively. 

\begin{figure}[t]
  \setlength{\unitlength}{1mm}
  \centering
 \includegraphics[width=\textwidth]{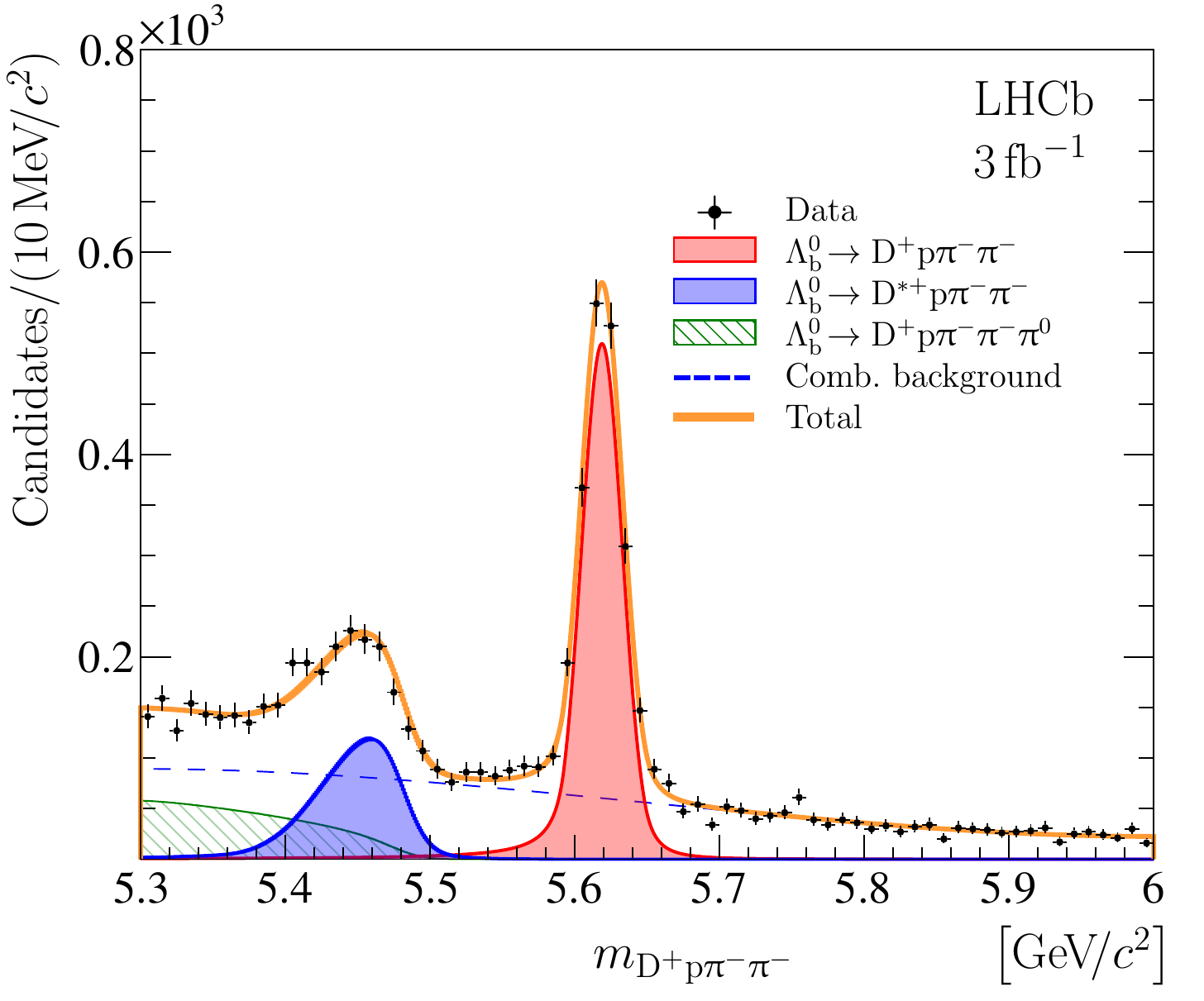} 
  \caption { \small
   Mass distribution for selected 
  \mbox{$\decay{\Lb}{\Dp\proton\pim\pim}$}~candidates.
   The~projection of an~unbinned likelihood fit, described in the~text, is superimposed.
  }
  \label{fig:signal}
\end{figure}

\begin{figure}[t]
  \setlength{\unitlength}{1mm}
  \centering
\includegraphics[width=\textwidth]{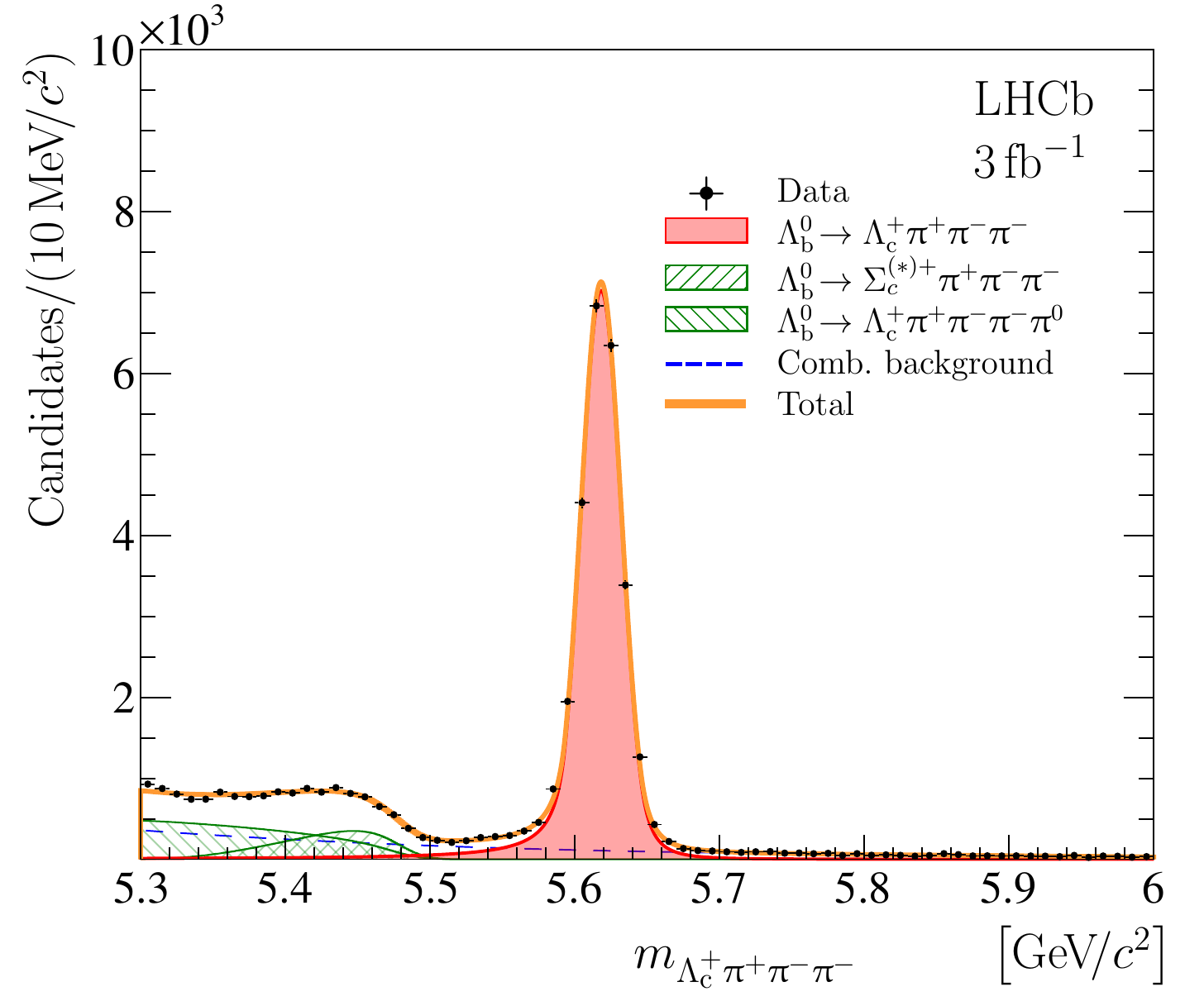}
  \caption{\small
  Mass distribution for selected 
  \mbox{$\decay{\Lb}{\Lc\pip\pim\pim}$}~candidates.
  The projection of an~unbinned likelihood fit, described in the~text, is superimposed.
  }
  \label{fig:normalisation}
\end{figure}

The reconstructed \Dp~and \Lc~candidates 
are combined with selected~$\proton\pim\pim$ and
$\pim\pip\pim$~candidates to form \Lb~candidates. 
Only~\Lb~candidates with a~transverse momentum 
above $3\gevc$ are selected for further analysis. 
To~improve the~mass resolution for the~$\Lb$\nobreakdash~candidates,
a~kinematic fit is performed~\cite{Hulsbergen:2005pu},
which constrains the~mass of the~$\Dp$ and $\Lc$~hadron candidates  
to their~known masses~\cite{PDG2021} 
and requires the~\Lb~candidate to originate 
from its associated PV.
A~requirement on the~\chisq from 
this fit further suppresses background.
The~reconstructed \Lb~decay vertex is required to be 
distinct from the PV, with the proper decay time 
of the~\Lb~candidate restricted to be 
above~$100\mum/c$.
The~proper decay time of the~\Dp and \Lc~candidates
calculated with respect to the~reconstructed 
\Lb~decay vertex is required
to be positive within the~resolution.
These two~requirements reduce 
the~background contributions 
from charmed hadrons produced directly
in the~\proton\proton~interaction, 
and random combinations of tracks
forming fake~\Dp~or \Lc~candidates.
At~least one track from the~selected \Lb~candidate 
must be matched with a~high energy deposit
in the~calorimeter system, used in the~hardware-trigger stage.
The~mass distributions for selected
\mbox{$\decay{\Lb}{\Dp\proton\pim\pim}$} and \mbox{$\decay{\Lb}{\Lc\pip\pim\pim}$}
candidates are shown in Figs.~\ref{fig:signal} and~\ref{fig:normalisation}, respectively. 


%

\section{Signal determination}
\label{sec:signals}
The~$\Dp\proton\pim\pim$~mass distribution shown in Fig.~\ref{fig:signal}
exhibits 
a~narrow peak corresponding to 
the~\mbox{$\decay{\Lb}{\Dp\proton\pim\pim}$}~decay.
In~addition, 
a~structure around \mbox{$5.4-5.5\gevcc$} is visible.
This~structure corresponds 
to~the~\mbox{$\decay{\Lb}{\Dstarp\proton\pim\pim}$}~decay
followed by the~decay of the~\Dstarp~meson
into $\Dp\piz$ or $\Dp\g$~states,
where the~neutral particle is not reconstructed. 
An~extended unbinned maximum\nobreakdash-likelihood fit 
to the~\mbox{${\Dp\proton\pim\pim}$}~mass distribution
is performed using  a~function 
consisting of a sum of the four following contributions.
\begin{itemize}
    \item A \mbox{$\decay{\Lb}{\Dp\proton\pim\pim}$}~component, 
    parameterised by a~modified Gaussian function
    with power\nobreakdash-law tails on both sides of 
    the distribution~\cite{Skwarnicki:1986xj,LHCb-PAPER-2011-013}.
    The~tail parameters are fixed to values 
    obtained from simulation, while the~width and 
    peak position  are allowed to vary in the~fit. 
    \item A \mbox{$\decay{\Lb}{\Dstarp\proton\pim\pim}$}~component,
    followed by \mbox{$\decay{\Dstarp}{\Dp\piz}$} and 
    \mbox{$\decay{\Dstarp}{\Dp\g}$}~decays.
    The~shape of the~component 
    is taken from simulation and modified by 
    a~first order positive polynomial which accounts for 
    the~unknown \Lb~decay model.
    The~parameters of the~polynomial function are allowed to 
    vary in the~fit. 
    \item A \mbox{$\decay{\Lb}{\Dp\proton\pim\pim\piz}$}~component,
    where the $\piz$~meson is undetected. 
    The~shape is also taken from simulation. 
    \item A combinatorial-background component,
    parameterised with a~positive monotonically-decreasing
     third\nobreakdash-order polynomial function.  
\end{itemize}
The~fit result is overlaid on Fig.~\ref{fig:signal}. 
The~signal yields for 
the \mbox{$\decay{\Lb}{\Dp\proton\pim\pim}$}
and \mbox{$\decay{\Lb}{\Dstarp\proton\pim\pim}$}~decays
are presented in Table~\ref{tab:yields}. 
A similar four-component 
function is used to describe the~$\Lc\pip\pip\pim$~mass spectrum.
\begin{itemize}
    \item A \mbox{$\decay{\Lb}{\Lc\pip\pim\pim}$}~component, 
    parameterised with a~modified Gaussian function
    with power\nobreakdash-law tails on both sides of 
    the distribution~\cite{Skwarnicki:1986xj,LHCb-PAPER-2011-013}.
    The~tail parameters are fixed to values 
    obtained from simulation, while the~width and 
    position are allowed to vary in the~fit. 
    \item A \mbox{$\decay{\Lb}{ 
      \Sigma_{\cquark}^{(\ast)+} \pip\pim\pim}$}~component, 
  followed by a~\mbox{$\decay{\Sigma_{\cquark}^{(\ast)+}}{\Lc\piz}$} decay with an~undetected \piz meson. 
   The~shape is taken from simulation. 
        \item A \mbox{$\decay{\Lb}{\Lc\pip\pim\pim\piz}$}~component,
    where the $\piz$~meson is undetected. 
    The~shape is taken from 
    simulation based on a~phase\nobreakdash-space  
    decay model. 
    \item A combinatorial-background component, parameterised with a~positive monotonically-decreasing
    third\nobreakdash-order polynomial function.  
\end{itemize}
The~fit result is overlaid on Fig.~\ref{fig:normalisation}
and the~signal yield for 
the~\mbox{$\decay{\Lb}{\Lc\pip\pim\pim}$}~decays 
is presented in Table~\ref{tab:yields}.

 \begin{table}[tb]
  \caption{\small 
 	Yields, \(N^{\mathrm{fit}}\), 
 	for the~\mbox{\(\decay{\Lb}{\Dp\proton\pim\pim}\)}, 
 	\mbox{\(\decay{\Lb}{\Dstarp\proton\pim\pim}\)} and
 	\mbox{\(\decay{\Lb}{\Lc\pip\pim\pim}\)}~decays
 	evaluated from fits to the \mbox{\(\Dp\proton\pim\pim\)} 
 	and \mbox{\(\Lc\pip\pim\pim\)}~mass spectra. 
 	The~yields with all corrections 
 	described in the~text, \(N^{\mathrm{cor}}\),  
 	are also given. 
 	The uncertainties are statistical only.
  }
  \label{tab:yields}
  \begin{center}
	\begin{tabular}{lr@{\,}c@{\,}lr@{\,}c@{\,}l}
    Decay mode 
    &  \multicolumn{3}{c}{\(N^{\mathrm{fit}}\ \ \ \ \)} 
    &  \multicolumn{3}{c}{\(N^{\mathrm{cor}}\ \ \ \ \)}
    \\[1.5mm]
    \hline 
    \\[-2mm]
      \(\decay{\Lb}{\Dp\proton\pim\pim}\)   
      & $1933$ & $\pm$ & $56$   
      & $1542$ & $\pm$ & $60$ 
      \\ 
      \(\decay{\Lb}{\Dstarp\proton\pim\pim}\)   
      & $862$ & $\pm$ & $55$    
      & $875$ & $\pm$ & $55$ 
     \\  
      \(\decay{\Lb}{\Lc\pip\pim\pim}\)  
      &  $ (26.51$ & $\pm$ & $0.18)\times 10^3$ 
      &  $ (25.91$ & $\pm$ & $0.18)\times 10^3$ 
   \end{tabular}
  \end{center}
  \end{table}

\begin{figure}[t]
\includegraphics[width=\textwidth]{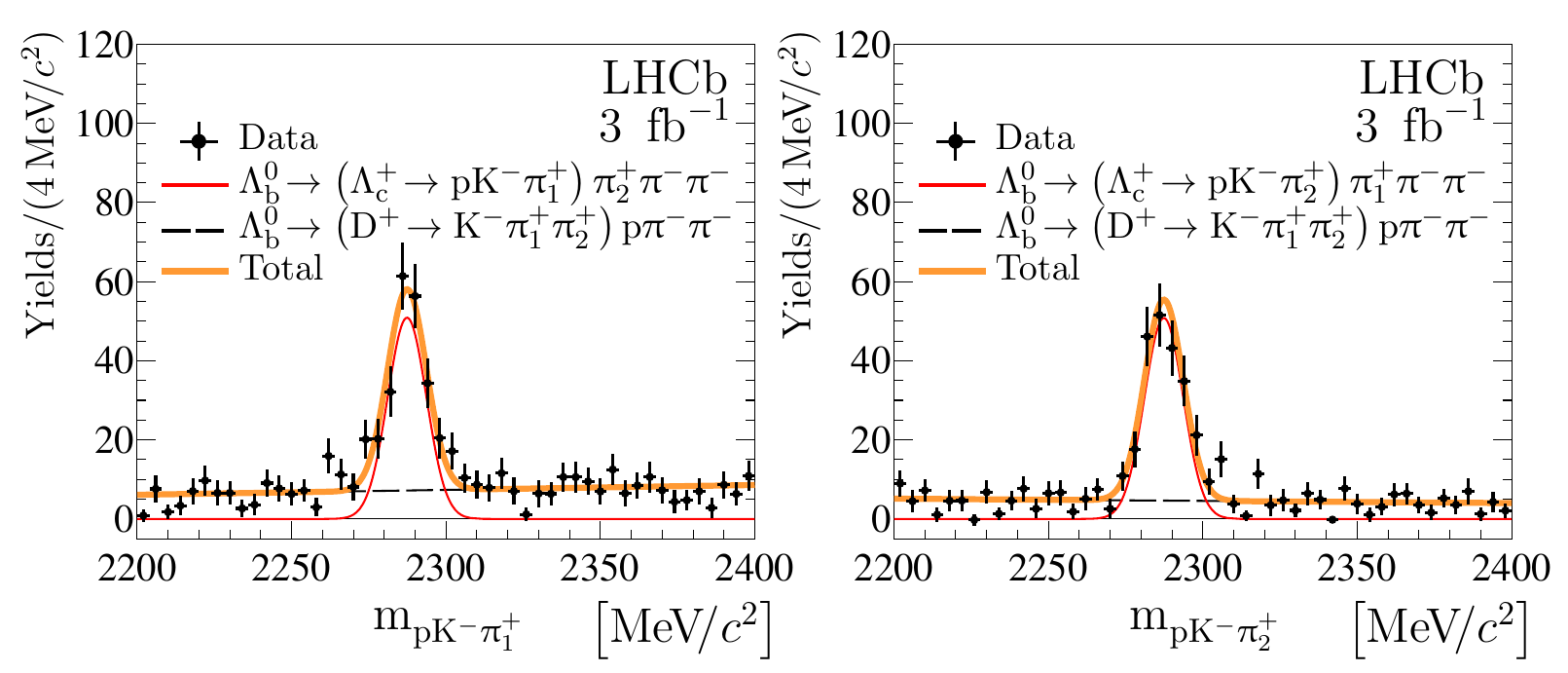}
  \caption{\small
  Background-subtracted
  $\proton\Km\Ppi^+_{1,2}$~mass spectra from 
  the~\mbox{$\decay{\Lb}{\Dp\proton\pim\pim}$}~channel
  with \mbox{$\decay{\Dp}{\Km\Ppi^+_1\Ppi^+_2}$} decays. 
  }
  \label{fig:crossfeed}
\end{figure}

Several corrections, described below, are applied
to the fitted yields.   
Since the \mbox{$\Dp\proton\pim\pim$} channel
with a \mbox{$\decay{\Dp}{\Km\pip\pip}$} decay
and the \mbox{$\Lc\pip\pim\pim$} channel 
with a~\mbox{$\decay{\Lc}{\proton\Km\pip}$} decay
consist of the~same final state particles, 
there can be cross\nobreakdash-feed between the two 
where true $\decay{\Lb}{\Lc\pip\pim\pim}$~decays 
are  misreconstructed as
\mbox{$\decay{\Lb}{\Dp\proton\pim\pim}$}~decays. 
This contribution is studied  using 
background\nobreakdash-subtracted 
$\proton\Km\pip$~mass distributions
from $\decay{\Lb}{\Dp\proton\pim\pim}$~decays.
The~\sPlot~technique~\cite{Pivk:2004ty} is applied 
to the~result of the~fit described above, 
using the~$\Dp\proton\pim\pim$~mass as the discriminating
variable.
The~resulting 
background\nobreakdash-subtracted
  $\proton\Km\Ppi^+_{1,2}$~mass spectra from 
  the~\mbox{$\decay{\Lb}{\Dp\proton\pim\pim}$}~channel
  with \mbox{$\decay{\Dp}{\Km\Ppi^+_1\Ppi^+_2}$} decays
are shown in Fig.~\ref{fig:crossfeed}.
The~peaks at the known mass of the~\Lc~baryon 
correspond to true
\mbox{$\decay{\Lb}{\Lc\pip\pim\pim}$}~decays
reconstructed as \mbox{$\decay{\Lb}{\Dp\proton\pim\pim}$}~decays.
Fits are performed to these distributions
with a~function consisting of the following two terms.
\begin{itemize}
\item The~\mbox{$\decay{\Lb}{\left( \decay{\Lc}{\proton\Km\pip} \right) \pip\pim\pim} $}~contribution
is modelled by a~Gaussian function with the~mean value and width taken
from the~fit to 
the~\mbox{$\decay{\Lc}{\proton\Km\pip}$}~signal 
from \mbox{$\decay{\Lb}{\left(\decay{\Lc}
{\proton\Km\pip}\right)\pip\pim\pim}$}~decays.
\item 
A~first\nobreakdash-order polynomial term models 
the~baseline background from \Lb~baryon 
decays without a $\Lc$~baryon 
in the~final state. 
\end{itemize} 
This fit yields~$395\pm 23$~misreconstructed 
\mbox{$\decay{\Lb}
{\left( \decay{\Dp}
{\Km\pip\pip} \right)\proton\pim\pim} $} candidates, 
which are subtracted
from the~total fit yield of
the~\mbox{$\decay{\Lb}{\Dp\proton\pim\pim} $} mode.
No~analogous pattern is observed 
in the~background\nobreakdash-subtracted 
$\proton\Km\pip$~mass spectra from 
$\decay{\Lb}{\Dstarp\proton\pim\pim}$~decays. 
Possible cross-feed from 
\mbox{$\decay{\Lb}{\D^{(*)+}\proton\pim\pim}$}~decays
in the \mbox{$\decay{\Lb}
{\left( \decay{\Lc}
{\proton\Km\pip} \right) \pip\pim\pim} $}~normalisation signal 
is found to be negligible.

\begin{figure}[t]
\includegraphics[width=\textwidth]{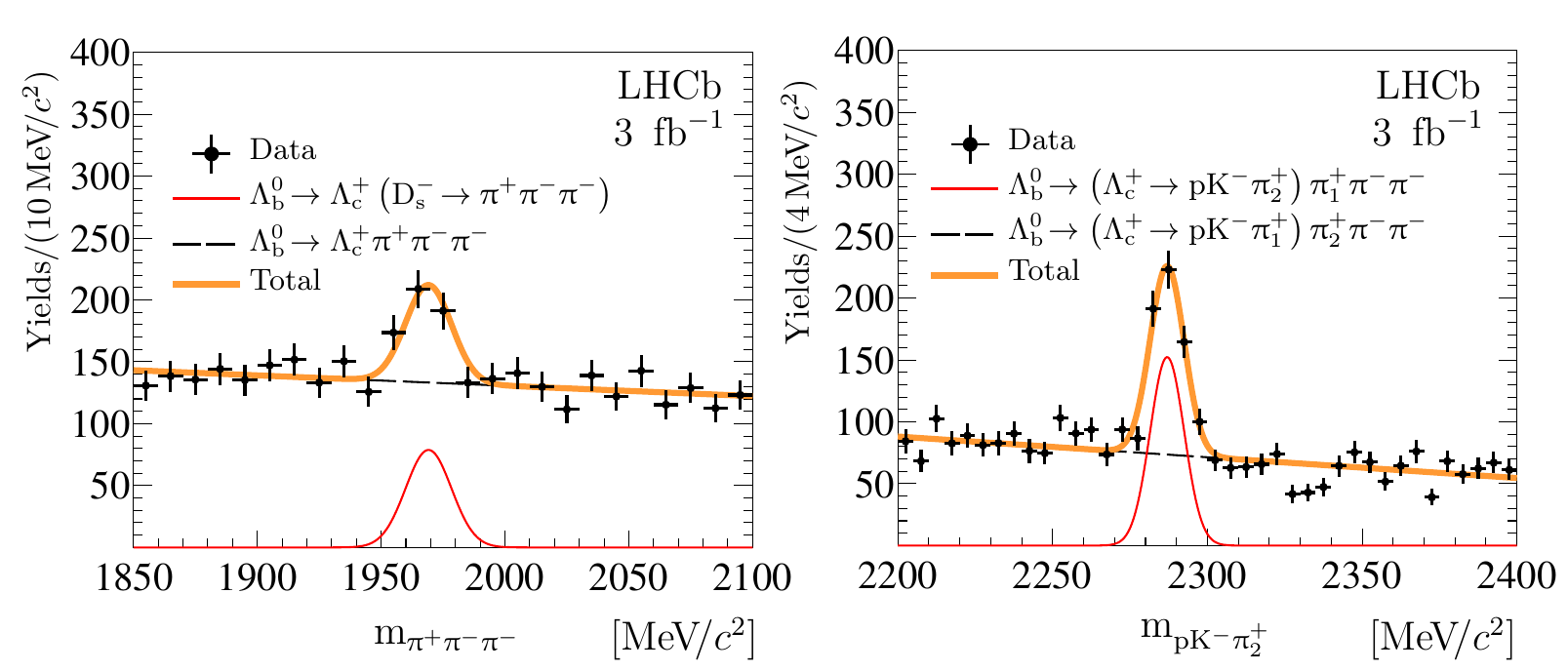}
  \caption{\small
  Background\protect\nobreakdash-subtracted (left)
  $\pip\pim\pim$~mass distribution
  from \mbox{$\decay{\Lb}{\Lc\pip\pim\pim}$} decays and (right) $\proton\Km\Ppi^+_2$~mass distribution
  from \mbox{$\decay{\Lb}
  {\left( \decay{\Lc}{\proton\Km\Ppi^+_1} \right)\Ppi^+_2\pim\pim }$}~decays.
  The results of the fits described in text are overlaid.
  }
  \label{fig:crossfeed2}
\end{figure}

The background-subtracted $\pip\pim\pim$~mass spectrum from 
\mbox{$\decay{\Lb}{\Lc\pip\pim\pim}$}~decays
is shown in Fig.~\ref{fig:crossfeed2} (left),
where background subtraction is performed with 
the \sPlot~technique using the \mbox{$\Lc\pip\pim\pim$}~mass 
as the~discriminating variable.
A~small contribution from 
\mbox{$\decay{\Lb}{\Lc\Dsm}$}~decays,
followed by 
the~\mbox{$\decay{\Dsm}{\pip\pim\pim}$}~decay,
is visible in the~$\pip\pim\pim$~mass 
spectrum~\cite{LHCb-PAPER-2014-002}. 
A~fit to the~background\nobreakdash-subtracted $\pip\pim\pim$~mass
distribution is performed   
with a~Gaussian function summed with 
a~positive first\nobreakdash-order 
polynomial function. The~Gaussian mean is set to the~known mass of the~\Dsm~meson~\cite{PDG2021}, 
while the~resolution is taken from
the~fit to 
a~larger data sample
available at an earlier stage of the~selection.
The~polynomial function models  
the~\mbox{$\decay{\Lb}{\Lc \pip\pim\pim}$}~decays
without an~intermediate $\Dsm$~meson.
The~fit yields \mbox{$176\pm 25$} \mbox{$\decay{\Lb}
{\Lc\Dsm}$} decays, which are subtracted  from 
the~total fitted yield of 
the~\mbox{$\decay{\Lb}{\Lc \pip\pim\pim}$} decays.

A small fraction of \mbox{$\decay{\Lb}
{ \left( \decay{\Lc}{\proton\Km\Ppi^+_1}
\right)\Ppi^+_2\pim\pim}$}~decays 
satisfies  all selection criteria after 
the~interchange of 
two~positive pions, \mbox{$\Ppi^+_1 \leftrightarrow \Ppi^+_2$}, 
causing  the~same six\nobreakdash-track 
combination to be reconstructed twice. 
This effect is studied using 
the~background\nobreakdash-subtracted 
$\proton\Km\Ppi^+_2$~mass distribution from 
the~selected \mbox{$\decay{\Lb}
{ \left( \decay{\Lc}{\proton\Km\Ppi^+_1}
\right)\Ppi^+_2\pim\pim}$}~decays,
shown in Fig.~\ref{fig:crossfeed2} (right). 
Duplicate candidates appear 
near the~known mass of the~\Lc~baryon.
A~fit to the~$\proton\Km\Ppi^+_{2}$~mass distribution 
is performed using a~Gaussian function for 
the~\mbox{$\decay{\Lb}
{ \left( \decay{\Lc}{\proton\Km\Ppi^+_2}
\right)\Ppi^+_1\pim\pim}$}~decays
and a first\nobreakdash-order polynomial function for the~\mbox{$\decay{\Lb}
{ \left( \decay{\Lc}{\proton\Km\Ppi^+_1}
\right)\Ppi^+_2\pim\pim}$}~decays.
The mean and width of the~Gaussian function are
taken from a fit to 
the~\mbox{$\decay{\Lc}{\proton\Km\pip}$} candidates
from \mbox{$\decay{\Lb}{ \left( 
\decay{\Lc}{\proton\Km\pip}\right) \pip\pim\pim}$}~decays.
The \mbox{$\decay{\Lb}
{ \left( \decay{\Lc}{\proton\Km\Ppi^+_2}
\right)\Ppi^+_1\pim\pim}$} yield is \mbox{$ 416\pm 32$}, 
and is subtracted from the~total fit yield of 
the~\mbox{$\decay{\Lb}{\Lc \pip\pim\pim}$}~decays.

Possible biases in 
the~yields of 
the~\mbox{$\decay{\Lb}{\D^{(*)+}\proton\pim\pim} $} and 
\mbox{$\decay{\Lb}{\Lc\pip\pim\pim} $}~decays from the relevant fits
are studied using pseudoexperiments.
The~largest bias is found 
for the~yield of \mbox{$\decay{\Lb}
{\Dstarp\proton\pim\pim} $}~decays and it is 1.5\%.
For~the~\mbox{$\decay{\Lb}{\Dp\proton\pim\pim} $}
and \mbox{$\decay{\Lb}{\Lc\pip\pim\pim} $}~decays 
the~corresponding biases are much smaller.
The~final yields, after all corrections described 
above are applied, are given in Table~\ref{tab:yields} for 
the~\mbox{$\decay{\Lb}{\D^{(*)+}\proton\pim\pim} $} and 
\mbox{$\decay{\Lb}{\Lc\pip\pim\pim} $}~decays.

\section{Efficiency and ratios of branching fractions}
\label{sec:efficiency}

The ratios \(\mathcal{R}_{\Dp}\) and \(\mathcal{R}_{\Dstarp}\), 
defined by Eq.~\eqref{eq:R} are calculated as
  \begin{subequations}
  \label{eq:rd:effcorr} 
  \begingroup
  \allowdisplaybreaks
  \begin{align}
    \mathcal{R}_{\Dp} & =    
     \dfrac {N^{\mathrm{cor}}_{\decay{\Lb}{\Dp\proton\pim\pim}}}
            {N^{\mathrm{cor}}_{\decay{\Lb}{\Lc\pip\pim\pim}}}
     \times 
       \dfrac 
           {\varepsilon_{\decay{\Lb}{\Lc\pip\pim\pim}}} {\varepsilon_{\decay{\Lb}{\Dp\proton\pim\pim}}}  
           \label{eq:rdp:effcorr}
           \\
           \intertext{and}
    \mathcal{R}_{\Dstarp} &=  
     \dfrac{N^{\mathrm{cor}}_{\decay{\Lb}{\Dstarp\proton\pim\pim}}}
          {N^{\mathrm{cor}}_{\decay{\Lb}{\Dp\proton\pim\pim}}}
      \times 
      \dfrac
          {\varepsilon_{\decay{\Lb}{\Dp\proton\pim\pim}}}
          {\varepsilon_{\decay{\Lb}{\Dstarp\proton\pim\pim}}}
          \label{eq:rdstarp:effcorr}\,,
  \end{align}
  \endgroup
  \end{subequations}
  where $N^{\mathrm{cor}}_{X}$ is the corrected~signal yield for decay 
  mode~$X$, 
  as per Table~\ref{tab:yields}, and 
  $\varepsilon_{X}$ is the~corresponding efficiency. 
  This~efficiency is defined as a product of the detector acceptance 
  $\varepsilon^{\mathrm{acc}}$, 
  reconstruction and 
  selection efficiency $\varepsilon^{\mathrm{rec\&sel}}$, 
  efficiency of the~hardware stage of the~trigger 
  $\varepsilon^{\mathrm{trg}}$ 
  and the~hadron\nobreakdash-identification 
  efficiency $\varepsilon^{\mathrm{PID}}$,
  \begin{equation}
  \label{eq:eff:compose} 
    \varepsilon = 
    \varepsilon^{\mathrm{acc}} 
    \varepsilon^{\mathrm{rec\&sel}}  
    \varepsilon^{\mathrm{trg}} 
    \varepsilon^{\mathrm{PID}}\,,  
  \end{equation}
  where each subsequent efficiency is defined with respect to the~product of previous efficiencies. 
  The~detector acceptance, and reconstruction and selection efficiency,  
  are determined using the simulation samples described in Sec.~\ref{sec:Detector}. 
  The~reconstruction and selection efficiency is corrected 
  for a small difference in the track reconstruction efficiency 
  between 
  data and simulation~\cite{LHCb-DP-2013-002}.
  The~trigger efficiency is calculated  
  from single-particle hadron-trigger efficiencies, 
  which are determined separately 
  for protons, kaons and pions
  from a~large  
  \mbox{$\decay{\Lb}
  {\left(\decay{\Lc}{\proton\Km\pip}\right)\pim}$}~data sample. 
  The~hadron\nobreakdash-identification efficiency is a combination of single-particle identification 
  efficiencies for protons, kaons and pions determined with large calibration samples of 
  \mbox{$\decay{\Lc}{\proton\Km\pip}$}, 
  \mbox{$\decay{\Lambda}{\proton\pim}$}, 
  \mbox{$\decay{\Dstarp}{ \left(  \decay{\Dz}{\Km\pip}\right) \pip}$}, 
  \mbox{$\decay{\Ds}{ \left( \decay{\Pphi}{\Kp\Km} \right) \pip}$} 
  and \mbox{$\decay{\KS}{\pip\pim}$}~decays in data~\cite{LHCb-DP-2012-003}. 
   The~ratios of efficiencies are, 
  \begin{subequations}\label{eq:efftot}
  \begingroup
  \allowdisplaybreaks
    \begin{align}
    \dfrac{\varepsilon_{\decay{\Lb}{\Dp\proton\pim\pim}}}  
          {\varepsilon_{\decay{\Lb}{\Lc\pip\pim\pim}}}
          & =  
          1.11 \pm 0.01 
          \label{eq:efftot:rdp} 
          \\ 
          \intertext{and} 
    \dfrac{\varepsilon_{\decay{\Lb}{\Dstarp\proton\pim\pim}}}
          {\varepsilon_{\decay{\Lb}{\Dp\proton\pim\pim}}}
          &=  
          0.93 \pm 0.01 
          \label{eq:efftot:rdstarp}\,,
    \end{align}
    \endgroup
  \end{subequations}
  where the uncertainties arise from the~finite size of the simulation samples. 
  Using the~corrected yields from
  Table~\ref{tab:yields} and efficiencies 
  from Eq.~\eqref{eq:efftot}, the~ratios 
  $\mathcal{R}_{\Dp}$ and $\mathcal{R}_{\Dstarp}$
  are found to be 
  \begin{subequations}
  \begingroup
  \allowdisplaybreaks
   \begin{align}
   \mathcal{R}_{\Dp}     
   & = 
   \left( 5.35 \pm 0.21 \right)\,\%   
   \\
   \intertext{and} 
   \mathcal{R}_{\Dstarp}
   & =  
   \left( 61.3 \pm 4.3\right)\,\% \,, 
  \end{align}
  \endgroup
 \end{subequations}
  where the~uncertainties  are statistical only.
  Systematic uncertainties are discussed 
  in Sec.~\ref{sec:systematics}.

\begin{figure}[htb]
%
\includegraphics[width=\textwidth]{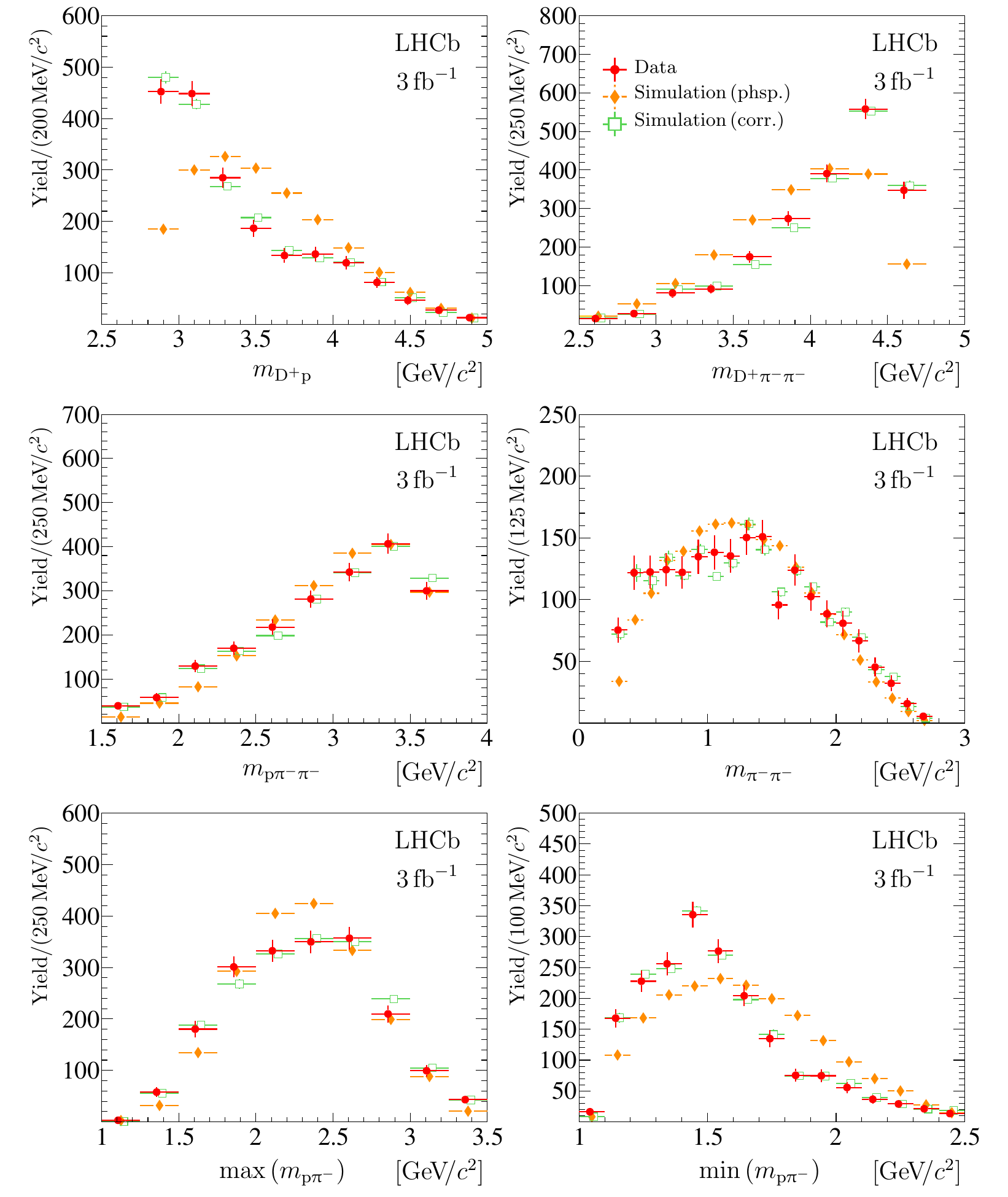}
\caption { \small
Background-subtracted 
$\Dp\proton$,
$\Dp\pim\pim$,
$\proton\pim\pim$,
$\pim\pim$,
and 
maximum and minimum $\proton\pim$~mass spectra 
for \mbox{$\decay{\Lb}{\Dp\proton\pim\pim}$}~decays.
Expectations from phase\protect\nobreakdash-space\,(phsp.) 
and corrected\,(corr.) 
simulation 
are overlaid.
}
  \label{fig:resonances_dplus}
 \end{figure}

\begin{figure}[t]
\includegraphics[width=\textwidth]{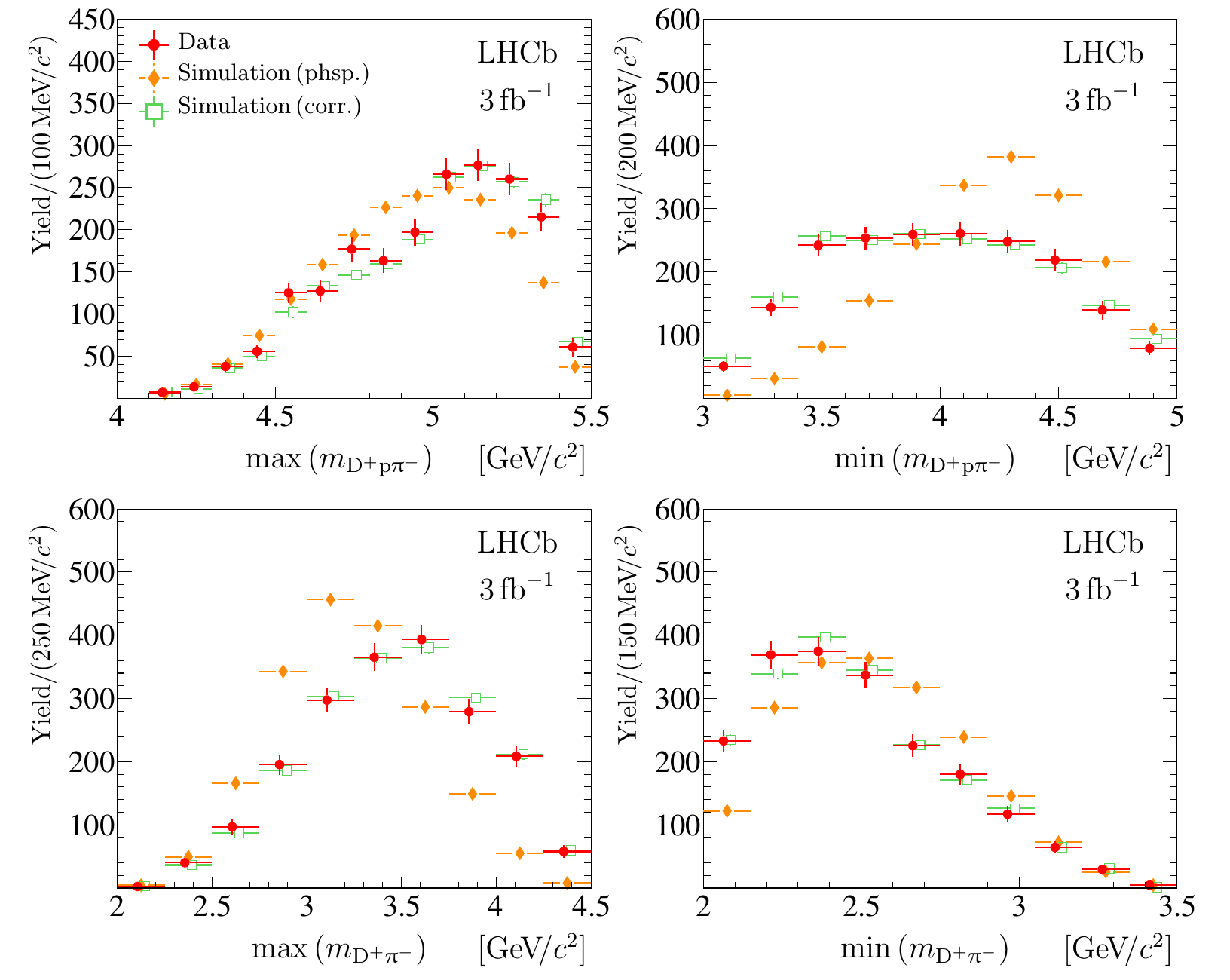}
\caption { \small
Background-subtracted 
maximum and minimum $\Dp\proton\pim$,
and maximum and  minimum $\Dp\pim$~mass 
spectra for 
\mbox{$\decay{\Lb}{\Dp\proton\pim\pim}$}~decays.
Expectations from phase\protect\nobreakdash-space\,(phsp.) 
and corrected\,(corr.) 
simulation 
are overlaid.
}
  \label{fig:resonances_dplus_rest}
  \end{figure}

\begin{figure}[t]
\includegraphics[width=\textwidth]{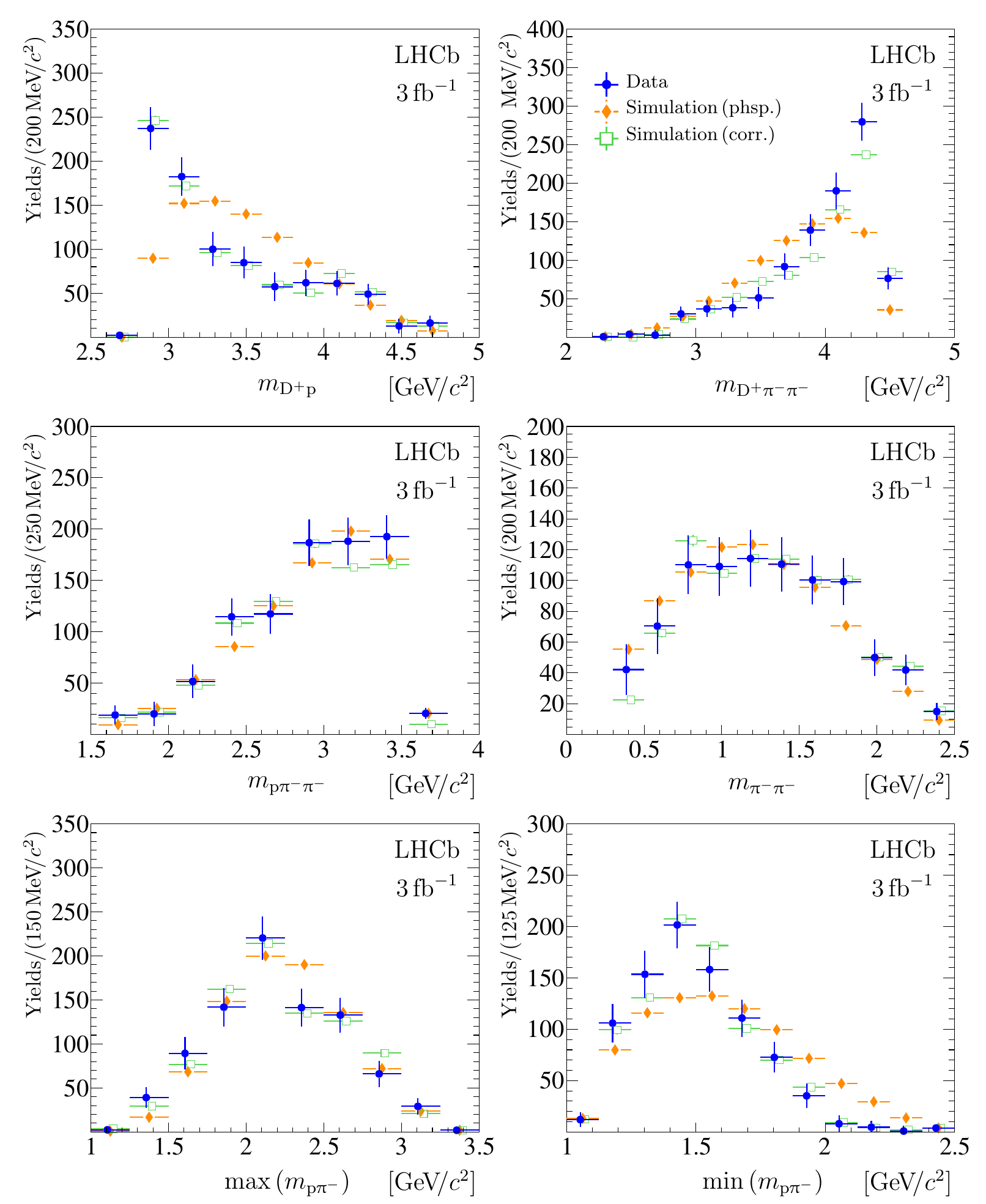}
\caption { \small
Background-subtracted
$\Dp\proton$,
$\Dp\pim\pim$,
$\proton\pim\pim$,
$\pim\pim$,
and 
maximum and minimum 
$\proton\pim$~mass spectra 
for 
\mbox{$\decay{\Lb}{\Dstarp\proton\pim\pim}$}~decays.
Expectations from phase\protect\nobreakdash-space\,(phsp.) 
and corrected\,(corr.) 
simulation 
are overlaid.
}
  \label{fig:resonances_dstar}
  \end{figure}

\begin{figure}[t]
\includegraphics[width=\textwidth]{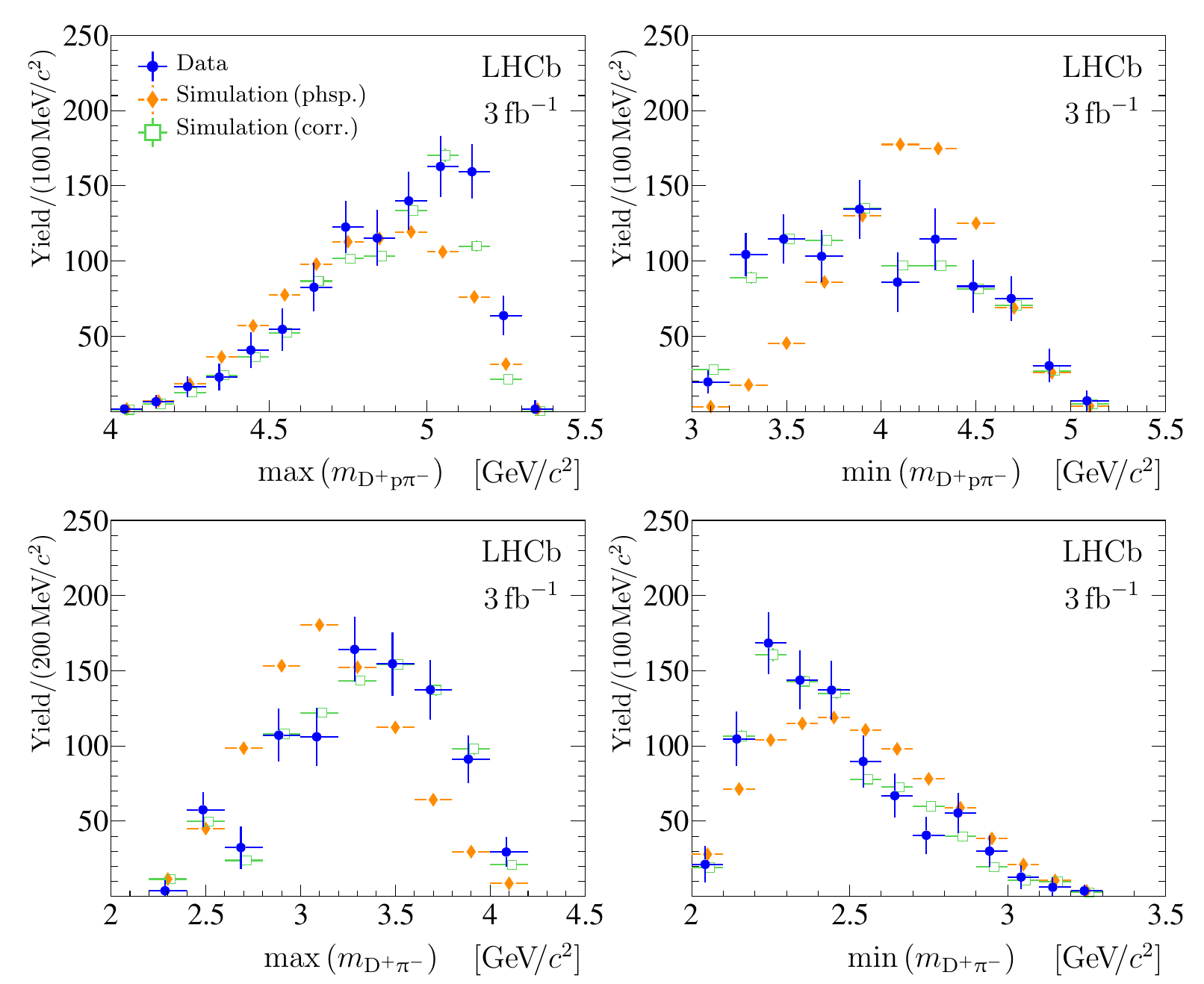}
\caption { \small
Background-subtracted 
maximum and minimum $\Dp\proton\pim$,
and maximum and minimum $\Dp\pim$~mass 
spectra for 
\mbox{$\decay{\Lb}{\Dstarp\proton\pim\pim}$}~decays.
Expectations from phase\protect\nobreakdash-space\,(phsp.) 
and corrected\,(corr.) 
simulation 
are overlaid.
}
  \label{fig:resonances_dstar_rest}
  \end{figure}

The~background-subtracted two- 
and three\nobreakdash-body mass spectra 
from the~\mbox{$\decay{\Lb}{\Dp\proton\pim\pim}$}
and \mbox{$\decay{\Lb}{\Dstarp\proton\pim\pim}$}~decays
are shown in Figs.~\ref{fig:resonances_dplus} 
through~\ref{fig:resonances_dstar_rest} 
with the~expectation 
from phase\nobreakdash-space 
simulated decays overlaid. 
The~\sPlot technique~\cite{Pivk:2004ty}
is used for background subtraction using
the~\Lb~candidate mass as 
a~discriminating variable.
The~analogous distributions for 
the~\mbox{$\decay{\Lb}{\Lc\pip\pim\pim}$}~decays
are shown in Appendix~\ref{sec:appendix};
corresponding distributions from 
the~corrected simulation samples, 
used for evaluation of 
the~efficiencies, are also shown. 
Large deviations between data and phase-space based simulation 
are observed, demonstrating a~rich structure of intermediate resonances
for the~decay of this study.

\def\myY{66}

\def\myY{66.}

\def\myY{66}

\afterpage{\clearpage} 

\section{Systematic uncertainties}
\label{sec:systematics}

Due to the shared analysis techniques used to determine 
the~yields for the~\mbox{$\decay{\Lb}{\D^{(*)+}\proton\pim\pim}$}
and \mbox{$\decay{\Lb}{\Lc\pip\pim\pim}$}~decays,
many systematic uncertainties cancel 
for the~ratios $\mathcal{R}_{\Dp}$
and $\mathcal{R}_{\Dstarp}$.
The~remaining contributions
to systematic uncertainty 
are summarised in Table~\ref{tab:systematic} 
and discussed below.

\begin{table}[b]
	\centering
	\caption{\small 
	Relative systematic uncertainties for 
	the~ratios $\mathcal{R}_{\Dp}$
	and  $\mathcal{R}_{\Dstarp}$.
	The total uncertainty is obtained by summing all 
	terms in quadrature.
	}
	\label{tab:systematic}
	\vspace{2mm}
	\begin{tabular*}{0.85\textwidth}{@{\hspace{3mm}}l@{\extracolsep{\fill}}cc@{\hspace{3mm}}}
	Source
    & $\upsigma_{\mathcal{R}_{\Dp}}~\left[ \%\right] $
    & $\upsigma_{\mathcal{R}_{\Dstarp}}~\left[ \%\right] $
    
   \\[1.5mm]
  \hline 
  \\[-2mm]
  Fit model                       &  1.5    &  5.7          
  \\  
  Multiple candidates  &  0.8    &  0.7          
  \\ 
  \Lb~kinematic spectra           
   &  0.2   &  0.4 
  \\  
  $\decay{\Lb}{\D^{(*)+}\proton\pim\pim}$~decay model 
  &  0.1  & 0.2   
  \\ 
  $\decay{\Lb}{\Lc\pip\pim\pim}$~decay model          
  &  0.3 &  ---  
  \\
  Hadron identification           &  0.7    & 0.5              
  \\
  Tracking efficiency             
  &  0.2    & 0.0 
  \\ 
  Hardware\nobreakdash-trigger efficiency     & 0.9    & 0.5   
  \\ 
  Data\nobreakdash-simulation difference      &  1.9    &  2.8          
  \\
  Simulation samples size        
  &   0.8  & 0.9
  \\[1.5mm]
  \hline 
  \\[-2mm]
  Total               &   2.9 &  6.5 
\end{tabular*}
	\vspace{3mm}
\end{table}

An~important source of systematic 
uncertainty on the ratios of the~branching fractions
arises from the~imperfect knowledge of the~mass 
shapes of 
the~signal and background components used in the~fits.
To~estimate this uncertainty,
several 
alternative models for the~signal and background components
are tested. 
For~the~\mbox{$\decay{\Lb}{\Dp\proton\pim\pim}$}
and \mbox{$\decay{\Lb}{\Lc\pip\pim\pim}$}~signal shapes 
the~tail parameters of modified Gaussian functions
are varied within uncertainties, determined
from fits to corresponding simulation samples. 
The~order of the positive monotonically\nobreakdash-decreasing 
polynomial function, used for 
modelling of the~background components, 
is varied between two and four.
The~ratio of branching  fractions 
for the~\mbox{$\decay{\Dstarp}{\Dp\g}$}
and \mbox{$\decay{\Dstarp}{\Dp\piz}$}~decays
affects the~shape of 
the~\mbox{$\decay{\Lb}{\Dstarp\proton\pim\pim}$}
component. This~ratio is varied within the~known 
uncertainty~\cite{CLEO:1992xqa,CLEO:1997rew,PDG2021}.
For the~\mbox{$\decay{\Lb}{\Dstarp\proton\pim\pim}$}~fit 
component, the polynomial factor 
that modifies the~shape obtained from the~simulation
is removed.
To~account for 
the~unknown resonance structure  for
the~\mbox{$\decay{\Lb}{\Dp\pip\pim\pim\piz}$}, 
\mbox{$\decay{\Lb}{\Sigma_{\cquark}^{(*)+}\pip\pim\pim}$}
and  
\mbox{$\decay{\Lb}{\Lc\pip\pim\pim\piz}$}~decays, 
the~corresponding fit components, determined from 
simulation, have been modified by the~positive\nobreakdash-definite 
linear polynomial functions.
The~parameters of these polynomial  functions 
are allowed to vary in the~fits.
For~each alternative model the~ratio of event yields
is determined, and the~maximal deviation
with respect to the~default model is taken as the systematic uncertainty. 
This~uncertainty is 1.5\% and~5.7\% for the~ratios $\mathcal{R}_{\Dp}$
and $\mathcal{R}_{\Dstar}$, respectively.

A~small fraction of  events contain 
multiple $\Lb$~candidates. 
These \Lb~candidates  have an approximately 
uniform mass distribution between 5.3 and 6.0\gevcc.   
To~estimate the~uncertainty associated with
the~presence of multiple \Lb~candidates, 
a single random \Lb~candidate is kept, 
while the other candidates are discarded
and the~ratios of the event yields are 
measured. This procedure is 
repeated for multiple trials
to mitigate the~effects of statistical 
fluctuations. The~differences between 
the~original
mean values of $\mathcal{R}_{\Dp}$ 
and $\mathcal{R}_{\Dstarp}$ and the values
obtained using randomisation 
are found to be
0.8\% and 0.7\%, respectively.
These differences 
are taken as systematic uncertainty
associated with 
the~selection of multiple candidates.

The~transverse momentum and rapidity  
spectra of \Lb~baryons in the simulation samples 
are corrected to reproduce those observed 
for the~\mbox{$\decay{\Lb}{\Lc\pip\pim\pim}$}~signal 
in data. This~correction is a~source of 
additional uncertainty,  which is evaluated with  
several sets of corrections obtained using 
different interval schemes for the~\(\pt\) and rapidity ~distributions of the \Lb candidates. 
These corrections are applied to the~simulation samples and maximal deviations 
of 0.2\% and 0.4\% are observed
for the~ratios $\mathcal{R}_{\Dp}$
and $\mathcal{R}_{\Dstar}$, respectively. 
These~deviations are set as the~systematic uncertainty
due to imperfect knowledge of the~production
spectra of the~\Lb~baryons.

 The~simulated 
 \mbox{$\decay{\Lb}{\Dp\proton\pim\pim}$},
 \mbox{$\decay{\Lb}{\Dstarp\proton\pim\pim}$} and 
 \mbox{$\decay{\Lb}{\Lc\pip\pim\pim}$}~samples
 are corrected to reproduce 
 the~two- and three\nobreakdash-body
 signal mass distributions observed in data. 
 Due to a large number of variables 
 and their correlations, 
 the~method requires several iterations to converge.
 The~corrections made for binned distributions 
 are illustrated in Figs.~\ref{fig:resonances_dplus} 
 through~\ref{fig:resonances_dstar_rest} 
 for the \mbox{$\decay{\Lb}{\Dp\proton\pim\pim}$},
 \mbox{$\decay{\Lb}{\Dstarp\proton\pim\pim}$} samples
 and Figs.~\ref{fig:resonances_norm_rest} and 
 \ref{fig:resonances_norm} 
 for the~\mbox{$\decay{\Lb}{\Lc\pip\pim\pim}$} sample.
 The~correction procedure 
 has been further validated by comparison
 of simulation and data for multiple 
 randomly constructed linear combinations of 
 the~ten 
 mass variables.
  To~estimate 
 the~systematic uncertainty related to the~imperfect
 knowledge of the~decay model for 
 \mbox{$\decay{\Lb}{\D^{(*)+}\proton\pim\pim}$} and 
 \mbox{$\decay{\Lb}{\Lc\pip\pim\pim}$}~decays, 
 the~number of~iterations is varied. 
 The~differences with respect to the~baseline results 
 for 
 the~$\mathcal{R}_{\Dp}$
and  $\mathcal{R}_{\Dstarp}$~ratios 
are assigned as systematic uncertainty 
due to the~imperfect knowledge of 
the~\mbox{$\decay{\Lb}{\D^{(*)+}\proton\pim\pim}$} and 
 \mbox{$\decay{\Lb}{\Lc\pip\pim\pim}$}~decay models.

 The~hadron\nobreakdash-identification 
 efficiency 
 for protons, kaons and pions 
 is estimated  using large calibration samples.
 The~uncertainty due to the~finite size
 of the~calibration samples is propagated to 
 the~ratios $\mathcal{R}_{\Dp}$
 and $\mathcal{R}_{\Dstarp}$ 
 using pseudoexperiments. 
 The~obtained variations 
 of 0.7\% and 0.5\% for
 the~$\mathcal{R}_{\Dp}$
and  $\mathcal{R}_{\Dstarp}$~ratios, respectively, 
are used as the systematic uncertainty
associated to the~hadron identification.

There are residual differences in 
the~reconstruction efficiency 
of charged\nobreakdash-particle tracks that 
do not cancel completely in the~ratio due 
to small differences in  
the kinematic distributions 
of the~final\nobreakdash-state particles.
The~track\nobreakdash-finding efficiencies 
obtained from 
simulation samples 
are corrected \mbox{using}
calibration modes~\cite{LHCb-DP-2013-002}.
The~uncertainties related to~the~efficiency 
correction factors are propagated to the~ratios of 
the~total efficiencies using pseudoexperiments
and are determined as 0.2\% and smaller than 0.1\%
for the~$\mathcal{R}_{\Dp}$
and $\mathcal{R}_{\Dstarp}$~ratios, respectively.
These values are taken as the systematic uncertainty 
associated with the tracking efficiency.

The~hardware\nobreakdash-trigger 
efficiency 
for protons, kaons and pions 
is estimated 
using a~large 
\mbox{$\decay{\Lb}{ \left( \decay{\Lc}{\proton\Km\pip}\right)\pim}$}~
calibration sample.
Efficiencies from alternative calibration samples, 
\eg \mbox{$\decay{\Dstarp}
{\left( \decay{\Dz}{\Km\pip} \right)\pip } $}~decays,
yield 0.9\% and 0.5\% variations 
for 
the~$\mathcal{R}_{\Dp}$
and $\mathcal{R}_{\Dstarp}$~ratios, respectively.
These~variations are 
taken as the~systematic uncertainty 
due to the~hardware\nobreakdash-trigger efficiency.

The~stability of the~results is checked by changing the~selection
criteria on transverse momenta for the~final 
state hadrons, 
the~$\chisq$ from the~kinematic fit
and decay time for \Lb~candidates. 
The~ratios~$\mathcal{R}_{\Dp}$
and $\mathcal{R}_{\Dstarp}$
vary by up to 1.9\% and 2.8\%, respectively, and 
these variation are
conservatively assigned as a~systematic uncertainty due to
data\nobreakdash-simulation differences
not considered elsewhere.
Finally, the~$0.8\%$ and 0.9\% relative uncertainties 
from Eq.~\eqref{eq:efftot} 
are assigned as a~systematic uncertainty 
due to the finite size of the simulated samples
for the~$\mathcal{R}_{\Dp}$
and $\mathcal{R}_{\Dstarp}$~ratios, respectively.
%

\section{Results and summary}
\label{sec:results}

The~decays \mbox{$\decay{\Lb}{\Dp\proton\pim\pim}$}
and \mbox{$\decay{\Lb}{\Dstarp\proton\pim\pim}$}
are observed using data collected with the LHCb detector
in proton\nobreakdash-proton collisions 
corresponding to 1~and~2\invfb
of integrated luminosity at 
centre\nobreakdash-of\nobreakdash-mass
energies of 7 and 8\tev, respectively. 
Both decay modes belong to the~relatively 
unexplored class of beauty\nobreakdash-baryon decays 
where the~\cquark-quark from the~$\bquark\to\cquark$~transition 
hadronises into the~final state separate from the~baryon,
\ie a~charm meson and a~proton. 
These multihadron decays exhibit a~rich resonance structure.

Using the~\mbox{$\decay{\Lb}{\Lc\pip\pim\pim}$}~decay as 
a~normalisation channel, the~ratios of branching fractions 
defined by Eq.~\eqref{eq:R} are measured to be
\begin{align*}
   \mathcal{R}_{\Dp}  
    & =  
    \left( 5.35 \pm 0.21 \pm 0.16 \right)\,\%  
    \\  
    \intertext{and} 
   \mathcal{R}_{\Dstarp} 
    & =    
   \left( 61.3 \pm 4.3 \pm 4.0 \right)\,\% \,, 
  \end{align*}
where the~first uncertainty 
is statistical and the~second systematic.
Using known branching fractions
for the~\mbox{$\decay{\Dp}{\Km\pip\pip}$} and 
\mbox{$\decay{\Lc}{\proton\Km\pip}$}~decays~\cite{PDG2021}
the~ratio of branching fractions for 
the~\mbox{$\decay{\Lb}{\Dp\proton\pim\pim}$}
and \mbox{$\decay{\Lb}{\Lc\pip\pim\pim}$}~decays
is found to be
\begin{equation*}
\dfrac{  \BR  ( \decay{\Lb}{\Dp\proton\pim\pim}  ) }
{ \BR (  \decay{\Lb}{\Lc\pip\pim\pim} ) } = \left( 3.58 \pm 0.14 \pm 0.11 \pm 0.19\right)\, \% \,,
\end{equation*}
where the~last uncertainty is 
due to imprecise knowledge of 
the~branching fractions
for the~\Lc~and \Dp~hadrons.

The~relative rate
for \mbox{$\decay{\Lb}{\Dstarp\proton\pip\pip}$}
and \mbox{$\decay{\Lb}
{\Dp\proton\pip\pip}$}~decays $r_{\Dstar}$ is defined as
\begin{equation*}
 r_{\Dstarp} \equiv 
 \dfrac{ \BR({\decay{\Lb}{\Dstarp\proton\pip\pip}} ) } 
       { \BR({\decay{\Lb}{\Dp\proton\pip\pip}}) }
 = 
 \dfrac{\mathcal{R}_{\Dstarp}}
 {\BR({\decay{\Dstarp}{\Dp\piz/\g}})}\,.
\end{equation*}
Using the~known branching 
fractions of the~\Dstarp~meson~\cite{PDG2021}, 
the~ratio $r_{\Dstarp}$ 
is~\mbox{$1.90\pm 0.19$}.
For~multihadron $\decay{\bquark}{\cquark}$~decays 
with a~large energy release, 
a~relative yield  of 
the~$\Dstarp$ and $\Dp$~mesons 
is expected to be similar to 
one for the~$\Dstarp$ and $\Dp$~ mesons  
produced via 
a~charm quark
fragmentation
in high\nobreakdash-energy hadron 
or $\epem$~interactions.   
A~na\"ive spin\nobreakdash-counting
rule~\cite{Braaten:1994bz,Falk:1993rf}
predicts 
the~ratio $r_{\Dstarp}$~to be as large as 3. 
The~relative production 
of \Dstarp and \Dp~mesons
produced promptly in 
$\proton\proton$~collisions
at $\sqs=5$, 7 and 13\tev 
is estimated using
the~cross sections 
of directly produced $\Dstarp$ and $\Dp$~mesons,
$\upsigma^{\mathrm{direct}}_{\decay{\proton\proton}{\Dstarp X}}$
and $\upsigma^{\mathrm{direct}}_{\decay{\proton\proton}{\Dp X}}$, as 
\begin{equation*}
r_{\Dstarp}^\textrm{\proton\proton} \equiv 
\dfrac{\upsigma^{\mathrm{direct}}_{\decay{\proton\proton}{\Dstarp X}}}
{\upsigma^{\mathrm{direct}}_{\decay{\proton\proton}{\Dp X}}} \approx 
\dfrac{ \upsigma_{\decay{\proton\proton}{\Dstarp X}}}
{ \upsigma_{\decay{\proton\proton}{\Dp X}} - 
\BR({ \decay{\Dstarp} {\Dp\piz/\g}})
\times\upsigma_{\decay{\proton\proton}{\Dstarp X}}}\,,
\end{equation*}
where $\upsigma_{\decay{\proton\proton}
{\Dstarp X}}$  and 
$\upsigma_{\decay{\proton\proton}{\Dp X}}$ 
are the~measured 
inclusive cross sections 
of  the promptly produced $\Dstarp$ and $\Dp$ mesons.
Assuming  an~independent fragmentation of 
the $\cquark$~quark 
into \Dstarp and \Dp~mesons in direct production, and 
averaging $r_{\Dstarp}^\mathrm{\proton\proton}$ 
over the different proton 
collision energies of 
5, 7, and 13\tev~\cite{LHCb-PAPER-2012-041, 
LHCb-PAPER-2015-041, 
LHCb-PAPER-2016-042}, 
the~value 
$r_{\Dstarp}^\textrm{\proton\proton}$ 
is $1.5\pm 0.1$. 
The~obtained  value 
is smaller than the~value of $r_{\Dstarp}$ obtained 
from the~\mbox{$\decay{\Lb}
{\D^{(*)+}}\proton\pim\pim$}~decays,
but consistent within two standard deviations.
The~value 
for the~ratio of production 
cross\nobreakdash-sections
of \Dstarp and  \Dp~mesons 
in $\epem$~collisions,  
$r^{\epem}_{\Dstar}=1.86\pm0.16$, 
from 
Ref.~\cite{Falk:1993rf}
is obtained from a combination of measurements
performed 
by the~CLEO~\cite{CLEO:1988jcc}, 
ARGUS~\cite{ARGUS:1991vjh}, 
ALEPH~\cite{ALEPH:1991phy}
and VENUS~\cite{VENUS:1993rpb}
collaborations
analysing data from high energy $\epem$~annihilation.
The~similarity between these values 
indicates a~possible correspondence between 
direct charm-meson production  
and fragmentation, and charm-meson production
in the~multihadron  decays of beauty hadrons.

Analysis of the~$\Lc\pip\pim\pim$~spectra shows that 
$\decay{\Lb}
{\Sigma_{\cquark}^{(*)+}\pip\pim \pim}$~decays  are 
largely suppressed with respect 
to $\decay{\Lb}{\Lc\pip\pim \pim}$~decays, 
see Fig.~\ref{fig:normalisation}.
The~relative production of charmed 
$\Sigma_{\cquark}^{(*)+}$ 
and $\Lc$~baryons exhibits the~same trend
both in $\epem$~annihilation~\cite{Belle:2017caf} 
and in high energy hadroproduction~\cite{ALICE:2021rzj}. 
From these measurements, 
a consistent picture emerges where formation 
and production of a~light isoscalar diquark 
that is a~scalar 
is more favourable 
during the~hadronisation of heavy charm quarks,
than a~light isovector diquark 
that is an~axial vector~\cite{Jaffe:2004ph,
Wilczek:2004im,
Selem:2006nd}. 
This~observation supports 
the~diquark model for heavy\nobreakdash-flavor 
baryon structure and production~\cite{Andersson:1983ia}.

In conclusion, \mbox{$\decay{\Lb}{\Dp\proton\pim\pim}$}
and \mbox{$\decay{\Lb}{\Dstarp\proton\pim\pim}$} decays
are observed for the~first time 
and their relative branching 
ratios are measured.
Both these decays, and 
the~\mbox{$\decay{\Lb}{\Lc\pip\pim\pim}$}~decays
used in this analysis as a~normalisation channel, 
demonstrate a~rich resonance structure. 
A~similarity  between prompt charm-meson production  
and charm-meson production from multihadron  
decays of \Lb baryons is observed. 
In~the~future, 
the~observed decay 
$\decay{\Lb}{\Dp\proton\pim\pim}$ can serve 
as a~normalisation mode for studies of 
similar rare decays, \eg
$\decay{\Xibz}{\Dp\proton\Km\pim}$ 
and 
$\decay{\Xibz}{\Dstarp\proton\Km\pim}$~decays.

\section*{Acknowledgements}
%
%
\noindent We express our gratitude to our colleagues in the~CERN
accelerator departments for the excellent performance of the~LHC. 
We~thank the~technical and administrative staff at the~LHCb
institutes.
We~acknowledge support from CERN and from the national agencies:
CAPES, CNPq, FAPERJ and FINEP,(Brazil); 
MOST and NSFC\,(China); 
CNRS/IN2P3\,(France); 
BMBF, DFG and MPG\,(Germany); 
INFN\,(Italy); 
NWO\,(Netherlands); 
MNiSW and NCN\,(Poland); 
MEN/IFA\,(Romania); 
MSHE\,(Russia); 
MICINN\,(Spain); 
SNSF and SER\,(Switzerland); 
NASU\,(Ukraine); 
STFC\,(United Kingdom); 
DOE NP and NSF\,(USA).
We~acknowledge the~computing resources that are provided by CERN, 
IN2P3\,(France), 
KIT and DESY\,(Germany), 
INFN\,(Italy), 
SURF\,(Netherlands),
PIC\,(Spain), 
GridPP\,(United Kingdom), 
RRCKI and Yandex LLC\,(Russia), 
CSCS\,(Switzerland), 
IFIN\nobreakdash-HH\,(Romania), 
CBPF\,(Brazil),
PL\nobreakdash-GRID\,(Poland) and 
NERSC\,(USA).
We~are indebted to the~communities behind the~multiple 
open\nobreakdash-source
software packages on which we depend.
Individual groups or members have received support from
ARC and ARDC\,(Australia);
AvH Foundation\,(Germany);
EPLANET, Marie Sk\l{}odowska\nobreakdash-Curie Actions and ERC\,(European Union);
A*MIDEX, ANR, IPhU and Labex P2IO, and
R\'{e}gion 
Auvergne\nobreakdash-Rh\^{o}ne\nobreakdash-Alpes\,(France);
Key Research Program of Frontier Sciences of CAS, CAS PIFI, CAS CCEPP, 
Fundamental Research Funds for the~Central Universities, 
and Sci. \& Tech. Program of Guangzhou\,(China);
RFBR, RSF and Yandex LLC\,(Russia);
GVA, XuntaGal and GENCAT\,(Spain);
the~Leverhulme Trust, 
the~Royal Society
 and UKRI\,(United Kingdom).

\clearpage
\appendix 

\renewcommand{\thetable}{A\arabic{table}}
 \renewcommand{\thefigure}{A\arabic{figure}}
 \renewcommand{\theequation}{A\arabic{equation}}
 \setcounter{figure}{0}
 \setcounter{table}{0}
 \setcounter{equation}{0}

\section{Mass spectra for \mbox{$\decay{\Lb}{\Lc\pip\pim\pim}$}~decays}
\label{sec:appendix}

\begin{figure}[htb]
\includegraphics[width=\textwidth]{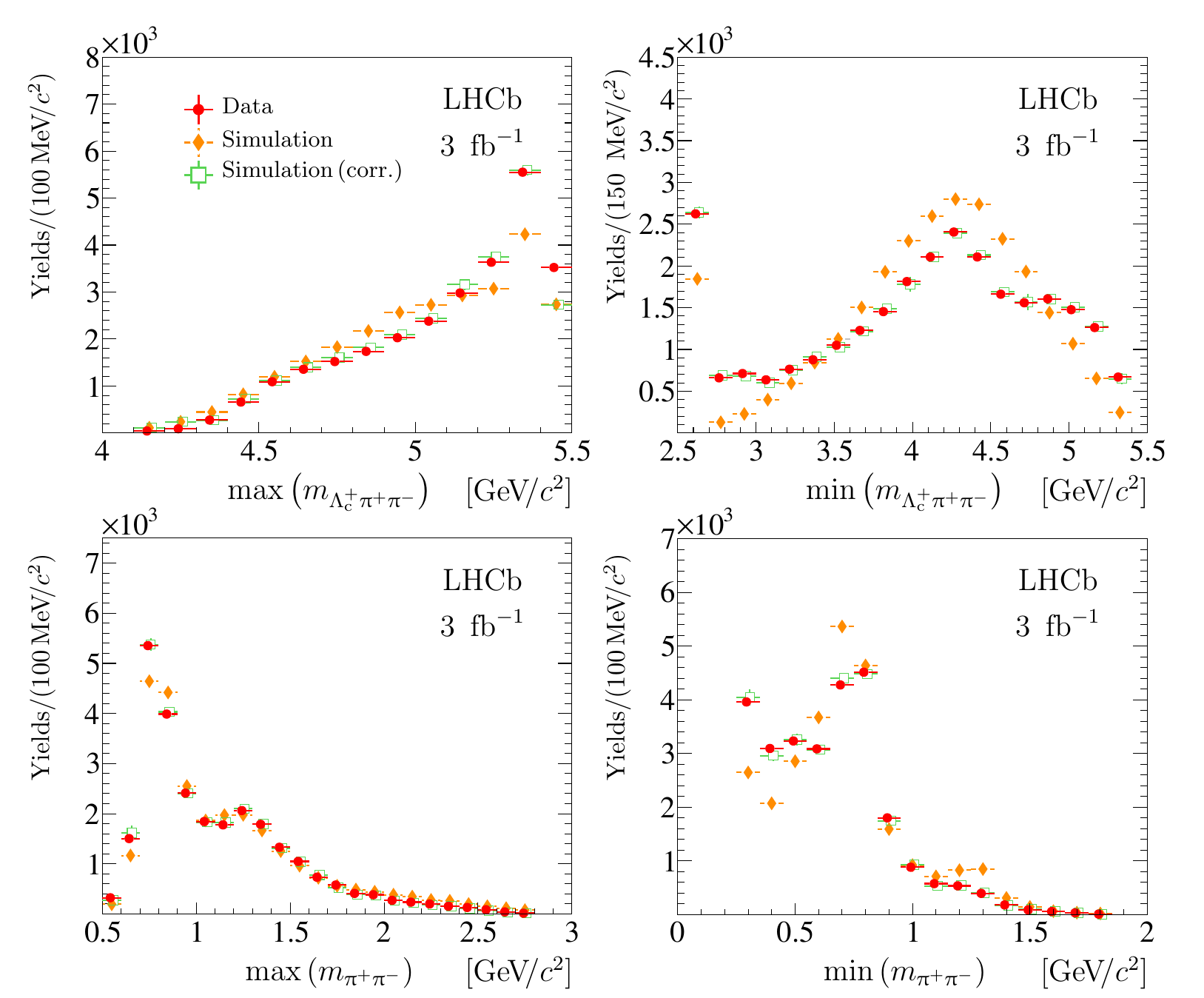}
  \caption { \small
Background-subtracted 
maximum and minimum $\Lc\pip\pim$, 
and maximum and minimum $\pip\pim$~mass 
spectra for 
the~\mbox{$\decay{\Lb}{\Lc\pip\pim\pim}$}~decays.
Expectations from uncorrected and corrected\,(corr.) simulation 
are overlaid. 
}
\label{fig:resonances_norm_rest}
 \end{figure}

\begin{figure}[htb]
\includegraphics[width=\textwidth]{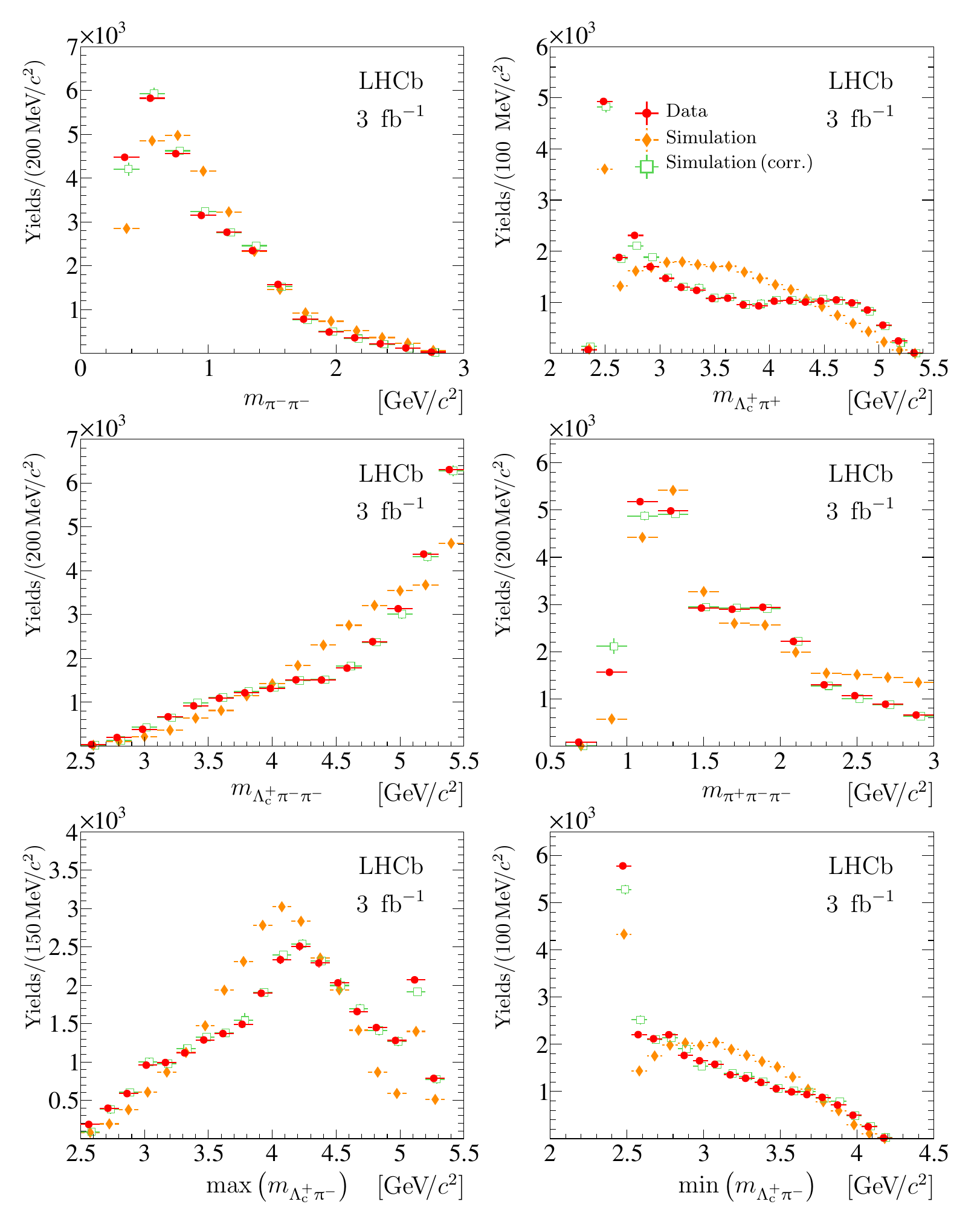}
  \caption { \small
 Background-subtracted 
 $\pim\pim$,
 $\Lc\pip$,
 $\Lc\pim\pim$,
 $\Lc\pip\pim\pim$,
 and maximum and  minimum $\Lc\pim$~mass 
 spectra for 
the~\mbox{$\decay{\Lb}{\Lc\pip\pim\pim}$}~decays.
Expectations from uncorrected and corrected\,(corr.) simulation 
are overlaid. 
  }
 \label{fig:resonances_norm}
 \end{figure}

Background\nobreakdash-subtracted 
two- and three-body  
mass spectra for 
the~\mbox{$\decay{\Lb}{\Lc\pip\pim\pim}$}~decays
are shown in Figs.~\ref{fig:resonances_norm_rest}
and~\ref{fig:resonances_norm}. 
A~rich structure of intermediate resonances
is visible.

\def\myY{66}

\def\myY{66}

\clearpage 
\addcontentsline{toc}{section}{References}
\bibliographystyle{LHCb}
\bibliography{main,standard,LHCb-PAPER,LHCb-CONF,LHCb-DP,LHCb-TDR}

\newpage
\centerline
{\large\bf LHCb collaboration}
\begin
{flushleft}
\small
R.~Aaij$^{32}$,
A.S.W.~Abdelmotteleb$^{56}$,
C.~Abell{\'a}n~Beteta$^{50}$,
F.~Abudin{\'e}n$^{56}$,
T.~Ackernley$^{60}$,
B.~Adeva$^{46}$,
M.~Adinolfi$^{54}$,
H.~Afsharnia$^{9}$,
C.~Agapopoulou$^{13}$,
C.A.~Aidala$^{87}$
% ...

{\footnotesize \it

% ...
$^{9}$Universit{\'e} Clermont Auvergne, CNRS/IN2P3, LPC, Clermont-Ferrand, France\\
% ...
$^{13}$LPNHE, Sorbonne Universit{\'e}, Paris Diderot Sorbonne Paris Cit{\'e}, CNRS/IN2P3, Paris, France\\
% ...
$^{32}$Nikhef National Institute for Subatomic Physics, Amsterdam, Netherlands\\
% ...
$^{46}$Instituto Galego de F{\'\i}sica de Altas Enerx{\'\i}as (IGFAE), Universidade de Santiago de Compostela, Santiago de Compostela, Spain\\
% ...
$^{50}$Physik-Institut, Universit{\"a}t Z{\"u}rich, Z{\"u}rich, Switzerland\\
% ...
$^{54}$H.H. Wills Physics Laboratory, University of Bristol, Bristol, United Kingdom\\
% ...
$^{56}$Department of Physics, University of Warwick, Coventry, United Kingdom\\
% ...
$^{60}$Oliver Lodge Laboratory, University of Liverpool, Liverpool, United Kingdom\\
% ...
$^{68}$Syracuse University, Syracuse, NY, United States\\
% ...
$^{87}$University of Michigan, Ann Arbor, United States, associated to $^{68}$\\
\bigskip
% ...
}
\end{flushleft}

\end{document}